\documentclass{aa}  

\usepackage{graphicx,lscape}
\usepackage[varg]{txfonts}

\usepackage{natbib}
\usepackage[english]{babel}

\usepackage{natbib,twoopt}

\newcommand{\ltapprox}{\raisebox{-0.5ex}{$\,\stackrel{<}{\scriptstyle\sim}\,$}}
\newcommand{\gtapprox}{\raisebox{-0.5ex}{$\,\stackrel{>}{\scriptstyle\sim}\,$}}
\newcommand{\zphot}{$z_{\rm phot}$~}

\newcommand{\Pint}{$P_{int}$~}

\usepackage{color}
\usepackage{amstext}

\begin{document}

   \title{The WIRCam Ultra Deep Survey (WUDS)}

   \subtitle{I. Survey overview and UV luminosity functions at z$\sim$5 and z$\sim$6 }


 \author{R.~Pell\'o\inst{\ref{inst1}}
          \and
P.~Hudelot\inst{\ref{inst2}}  \and
N.~Laporte\inst{\ref{inst3},\ref{inst1}} \and
Y.~Mellier\inst{\ref{inst2}}  \and
H. J.~McCracken\inst{\ref{inst2}}  \and
M.~Balcells\inst{\ref{inst4}}  \and
F.~Boone\inst{\ref{inst1}} \and
N.~Cardiel\inst{\ref{inst6}}  \and
J.~Gallego\inst{\ref{inst6}}  \and
F.~Garz\'on\inst{\ref{inst4},\ref{inst5}}  \and
R.~Guzm\'an\inst{\ref{inst7}}  \and
J.F.~Le Borgne\inst{\ref{inst1}} \and
M.~Prieto\inst{\ref{inst4},\ref{inst5}}  \and
J.~Richard\inst{\ref{inst8}}  \and
D.~Schaerer\inst{\ref{inst9},\ref{inst1}} \and
L.~Tresse\inst{\ref{inst8}} \and
S.~Arnouts\inst{\ref{inst10}}  \and
J.G.~Cuby\inst{\ref{inst10}}  \and
K.~Disseau\inst{\ref{inst8}}  \and
M.~Hayes\inst{\ref{inst11}} 
      }


\institute{
  Institut de Recherche en Astrophysique et Planétologie (IRAP), Université de Toulouse, CNRS, UPS, CNES, 14 Av. Edouard Belin, F-31400 Toulouse, France\email{rpello@irap.omp.eu}\label{inst1}
\and
Institut d'Astrophysique de Paris, UMR7095 CNRS, Universit\'e Pierre \&  Marie
Curie, 98 bis boulevard Arago, 75014 Paris, France\label{inst2}
\and
Department of Physics and Astronomy, University College London, Gower Street, London WC1E 6BT, UK\label{inst3}
\and
Instituto de Astrof\'isica de Canarias, c/Via Lactea s/n, 38205, La Laguna,
Tenerife, Spain\label{inst4}
\and
Departamento de Astrof\'isica, Universidad de La Laguna, Avda. Astrof\'isico Fco. Sánchez, E-38206 La Laguna, Tenerife, Spain\label{inst5}
\and
Departamento de Astrof\'isica y CC de la Atm\'osfera, Universidad Complutense
de Madrid, Av Complutense s/n, 28040 Madrid, Spain\label{inst6}
\and
Department of astronomy, University of Florida, 211 Bryant Space Science
Center, Gainsville, FL 32611-2055, USA\label{inst7}
\and
Univ Lyon, Univ Lyon1, Ens de Lyon, CNRS, Centre de Recherche Astrophysique de Lyon (CRAL) UMR5574, F-69230, Saint-
Genis-Laval, France\label{inst8}
\and
Geneva Observatory, 51, Ch. des Maillettes, CH-1290 Versoix, Switzerland\label{inst9}
\and
Laboratoire d'Astrophysique de Marseille, CNRS - Universit\'e Aix-Marseille,
38 rue Fr\'ed\'eric Joliot-Curie, 13388 Marseille Cedex 13, France\label{inst10}
\and
Stockholm University, Department of Astronomy and Oskar Klein Centre for Cosmoparticle Physics, AlbaNova University Centre, SE-10691, Stockholm, Sweden\label{inst11}
          }

   \date{Received ; accepted }

 
  \abstract
   {The aim of this paper is to introduce the WIRCam Ultra Deep Survey (WUDS), a near-IR
photometric survey carried out at the CFH Telescope in the field of the
CFHTLS-D3 field (Groth Strip).
WUDS includes four near-IR bands ($Y$, $J$, $H$ and $K_s$) over a field of
view of $\sim$400 arcmin$^2$.
The typical depth of WUDS data reaches between $\sim$26.8
in $Y$ and $J$, and $\sim$26 in $H$ and $K_s$ (AB, 3$\sigma$ in 1.3\arcsec aperture),
whereas the corresponding depth of the CFHTLS-D3 images in this region ranges between
28.6 and 29 in $ugr$, 28.2 in $i$ and 27.1 in $z$ (same S/N and aperture).
The area and depth of this survey were
specifically tailored to set strong constraints on the cosmic star formation rate and the
luminosity function brighter or around $L^{\star}$ in the z$\sim$6-10 redshift domain,
although these data are also useful for a variety of extragalactic projects. 

This first paper is intended to present the properties of the public WUDS survey in details:
catalog building, completeness and depth, number counts, photometric redshifts,
and global properties of the galaxy population. 
We have also concentrated on the selection and characterization of galaxy
samples at z$\sim [$4.5 -- 7$]$ in this field. For these purposes, we
include an adjacent shallower area of $\sim$1260 arcmin$^2$ in this region,
extracted from the WIRCam Deep Survey (WIRDS), and observed in $J$, $H$ and $K_s$ bands.
UV luminosity functions were derived at z$\sim$5 and z$\sim$6 taking advantage from the fact that 
WUDS covers a particularly interesting regime at intermediate luminosities,
which allows a combined determination of M$^{\star}$ and $\Phi^{\star}$ with increased 
accuracy.

Our results on the luminosity function are consistent with a small
evolution of both M$^{\star}$ and $\Phi^{\star}$
between z$=$5 and z$=$6, irrespective
of the method used to derive them, either photometric redshifts applied to
blindly-selected dropout samples or the classical Lyman Break Galaxy
color-preselected samples. Our results lend support to higher $\Phi^{\star}$
determinations at z$=$6 than usually reported. 
The selection and combined analysis of different galaxy samples at z$\ge$7
will be presented in a forthcoming paper, as well as the evolution of the UV
luminosity function between z$\sim$ 4.5 and 9. WUDS is intended to provide a
robust database in the near-IR for the selection of targets for detailed
spectroscopic studies, in particular for the EMIR/GTC GOYA Survey.
}


   \maketitle
%


\section{Introduction}

   This paper introduces the WIRCam Ultra Deep Survey (WUDS), a public near-IR
photometric survey carried out at the CFH Telescope in the field of the
CFHTLS-D3 field (Groth Strip). The area and depth of this survey were
specifically tailored to set strong constraints on the cosmic star-formation
rate (SFR) and the UV luminosity function (hereafter LF) around or brighter than $L^{\star}$ in the
z$\sim$6-10 redshift domain, taking advantage from the large field of view and
sensitivity of WIRCam. Determining the precise contribution of star-forming sources at z$\ge6$
to the cosmic reionization remains an important challenge for modern cosmology. The
study of their physical properties, starting with the spectroscopic
confirmation of current photometric candidates, requires the use of the most efficient
ground-based and space facilities presently available and, in practice, this exercise is
limited to the brightest candidates. Although the motivation of WUDS is clearly
focused on the high-z universe, these data are also useful for a variety of
extragalactic projects.

Deep and/or wide-field surveys in the near-IR bands are recognized since the
pioneering studies in the 90's as key observations to understand the process of
galaxy evolution at intermediate redshifts, in particular to address the process of
stellar mass assembly at 1$\le$z$\le$3 
\citep[see e.g.,][and the references therein]{1994ApJ...434..114C, 
2002A&A...381L..68C, 
2003AJ....125.1107L, 
2012A&A...544A.156M, 
2013ApJ...775..106C, 
2015ApJ...803...26P, 
2016ApJS..224...24L}. 
One of the main applications of deep near-IR photometry is the selection of
star-forming galaxies at z$\sim6-10$ based on their rest-frame UV continuum,
a study conducted during the last decade in a context of international competition
using the Hubble Space Telescope (HST) and various ground-based facilities, both in
blank and in lensing fields 
\citep[see e.g.,][and the references therein]{
2004ApJ...607..697K, 
2004A&A...416L..35P, 
2004ApJ...616L..79B, 
2010MNRAS.409..855B, 
2010ApJ...709L..16O, 
2012ApJ...756..164F, 
2013ApJ...762...32C, 
2013MNRAS.432.2696M, 
2015ApJ...803...34B, 
2018PASJ...70S..10O}. 
The efficiency on the selection of high-z galaxies depends on the availability of
ultra-deep multiwavelength data, using an appropriate set of near-IR filters in
combination with optical data. Although lensing clusters and ultra-deep pencil-beam
surveys are more efficient to conduct detailed studies toward the faint-end of the LF,
given the strong field to field variance in number counts in these regimes, 
observations of wide blank fields are mandatory and equally important to set
reliable constraints on the brightest end of the UV LF.
Figure~\ref{WUDS_surveys} displays a comparison between the effective area
versus depth for different representative ``deep'' NIR surveys available 
in the H-band ($\sim$5$\sigma$). WUDS covers an interesting niche between 
wide (but still deep) surveys such as UltraVISTA \citep{2012A&A...544A.156M, 
2016ApJS..224...24L}, 
and ultra-deep pencil beam surveys such as the eXtreme Deep Field
\citep[XDF,][]{2013ApJS..209....6I}, 
or lensing clusters, such as CLASH \citep{2012ApJS..199...25P} 
and the Hubble Frontier Fields \citep{2017ApJ...837...97L}. 
In this respect, the region covered by WUDS is a single field in the northern hemisphere,   
with an area comparable to the HST MCT CANDELS-N Survey 
\citep{2011ApJS..197...35G, 
2011ApJS..197...36K}. 
Contrary to CANDELS WFC3 imaging, which is limited to wavelengths up to 1.6$\mu$m
(excepted for some $K_s$-band imaging on the CANDELS-S),
WUDS also includes deep $K_s$-band imaging, and a complete and homogeneous coverage in the visible domain
from the CFHTLS-D3 field (Groth strip), that is a photometric catalog survey
in nine filter-bands ($ugrizYJHK_s$). 

WUDS was originally proposed by the Galaxy Origins
and Young Assembly (GOYA\footnote{\tt{http://www.astro.ufl.edu/GOYA/home.html}})
team as part of the effort
for the exploitation of the multiobject near-IR spectrograph EMIR at the GTC
\citep{Balcells03,2016SPIE.9908E..1JG}, 
in particular to provide a robust selection of targets for observations with EMIR.
EMIR is a wide-field, near-IR spectrograph commissioned in 2016 at the Nasmyth A
focus of the Spanish GTC at Canary
Islands\footnote{\url{http://www.gtc.iac.es/instruments/emir/emir.php}}.
EMIR is one of the first fully cryogenic multiobject spectrographs to be operated
on a 10m-class telescope, with a spectral resolution R$\sim$4000-5000, 
high enough to achieve an efficient OH-line suppression. It was specifically designed 
for the study of distant galaxies, in particular the GOYA project to be carried
as part of the EMIR's GTO program 
\citep[see e.g.,][]{Guzman03, Balcells03, Garzon07}.

In this paper we have concentrated on the presentation of the survey, as well as on 
the selection and characterization of galaxy samples at z$\sim [$4.5 -- 7$]$
in this field. The results obtained on the LF at z$\ge 7$, as well as the
evolution of the bright edge of the UV LF between z$\sim$ 4.5 and 9,
will be presented in a separate paper
(Laporte et al. 2018; hereafter Paper II).

In Sect.\ \ref{observations} we describe the WUDS observations. Sect.\ \ref{reduction}
is devoted to data processing and the construction of the image dataset.
Sect.\ \ref{catalog_built} presents the extraction of sources and the construction of photometric catalogs.
The characterization of the photometric survey is given in Sect.\ \ref{catalogs},
including the determination of the depth and completeness of the survey, 
and number counts. Sect.\ \ref{data_lowz} presents the quality achieved
in the computation of photometric redshifts up to z$\sim$1.5, and the global
properties of the galaxy population at low-$z$.
Sect.\ \ref{selection-results} presents the selection of candidates in the redshift interval
z$\sim [$4.5 -- 7$]$. The properties of these samples of galaxies are studied, in
particular the luminosity function (LF), and compared to previous findings.
Conclusions and perspectives are given in Sect.\ \ref{conclusion}.

   Throughout this paper, a concordance cosmology is adopted, with
$\Omega_{\Lambda}=0.7$, $\Omega_{m}=0.3$ and $H_{0}=70\ km\ s^{-1}\ Mpc^{-1}$.
All magnitudes are given in the AB system \citep{Oke-Gunn}. Table~\ref{tab_phot}
presents the conversion values between Vega and AB systems for
our photometric dataset.
Data products described in this paper are available through the following website: 
\url{http://wuds.irap.omp.eu/}

   \begin{figure}
   \centering
   \includegraphics[width=0.48\textwidth]{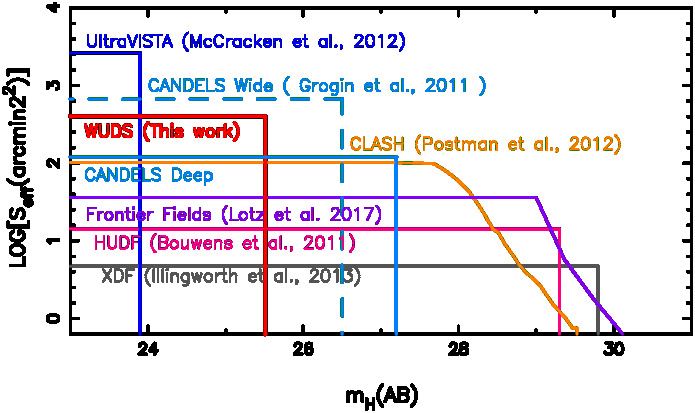}
   \caption{Effective area versus depth for different
NIR surveys available in the $H$-band ($\sim$5$\sigma$), both in blank fields
(UltraVISTA, CANDELS, HUDF and XDF)
and lensing clusters (CLASH and Hubble Frontier Fields). The appropriate
correction was applied to CLASH and Hubble Frontier Fields to account for lensing effects
on the effective area beyond the limiting magnitude,
assuming z$\sim$7 for the source plane. 
}
         \label{WUDS_surveys}
   \end{figure}
%
\section{Observations}
\label{observations}

   Observations were carried out with WIRCam at the Canada France Hawaii 
Telescope. WIRCam\footnote{\tt{http://www.cfht.hawaii.edu/Instruments/Imaging/WIRCam/}}
is a wide-field near-IR camera with four HAWAII2-RG detectors, 
$2048\times2048$ pixels each, and a pixel scale of 0.306\arcsec. The total
WIRCam field of view is 21.5\arcmin $\times$ 21.5\arcmin,
with a gap of 45\arcsec \ between adjacent detectors. 

\begin{figure}
   \centering
   \includegraphics[width=0.55\textwidth]{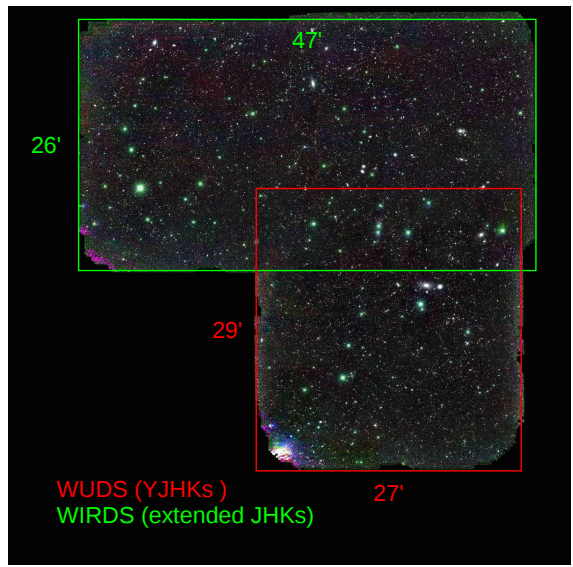}
      \caption{Layout of the WUDS Survey showing the regions covered by the
deep WUDS survey (red line) and the extended WIRDS area (green line) 
within the CFHTLS-D3 field (black area, $\sim$1 deg$^2$).
north is up and east is to the left.
}
         \label{WUDS_layout}
   \end{figure}

   WUDS was carried out on the CFHTLS-D3 (Groth Strip), on a field 
centered around $\alpha$=14:18:15.3 $\delta$=+52:38:45 (J2000), in order to
avoid the presence of bright stars while maximizing the coverage by other
surveys, in particular the DEEP Groth Strip Survey
\citep{2005ApJS..159...41V, 
2005ApJ...620..595W} 
and the AEGIS Survey\footnote{\tt{http://aegis.ucolick.org}}
\citep{Davis2007}. 
The WUDS pointing was chosen in such a way that observations could be combined
with public CFHTLS-Deep data on this field, obtained through five bands in the
optical domain, namely $u^{\ast}$, $g$, $r$, $i$, and $z$ (see
TERAPIX\footnote{http://terapix.iap.fr}).
Fig.\ \ref{WUDS_layout} presents the layout of WUDS showing
the regions covered by the different data sets on the CFHTLS-D3 field.

   WUDS images were obtained in queue scheduling mode between May and July 2008, 
in the four broad-band filters of WIRCam, namely $Y$, $J$, $H$ and $K_s$,
covering a field of view of $\sim$400 arcmin$^2$. This means a single shot
with WIRCam, with nodding and dithering configurations allowing us to
maximize the area covered with more than 75\% of the total exposure time
over the WIRCam field. In addition to these observations, we included 16.6h of
exposure time on the same area, obtained by the WIRDS Survey in $J$, $H$ and
$K_s$ \citep{Bielby2012}. Table~\ref{tab_wuds} summarizes these
observations, referenced as ``WUDS'' in Table~\ref{tab_phot}, also corresponding to
the area delimited in red in Figure~\ref{WUDS_layout}. 

   In order to obtain well sampled images given the goal seeing conditions, 
and to achieve an optimum matching with the CFHTLS-D3 images, on-target
observations were performed using the micro-dithering pattern of WIRCam. 
This consists of a $2\times2$ dithering pattern with offset positions
separated by 0.5 pixels, constituting a ``data-cube''. 
Each data-cube in Tab.\ \ref{tab_wuds} contains four
such images. Table~\ref{tab_wuds} also reports the individual exposure times
and the number of exposures needed to complete the total exposures.
Observations were performed with large dithering patterns (the
equivalent of 1/2 of detector, the four detectors being separated by 45\arcsec; see above)
and large overlaps providing an optimized gap filling and also a
better object removal for sky-subtraction (see below).
Exposure times in the near-IR were setup in such a way that a
good S/N is achieved for the detection of the rest-frame UV continuum of
Lyman-Break Galaxy samples (hereafter LBG) given the depth of the CFHTLS-D3
images in the optical domain (see also Sect.\ \ref{selection}). 
We have also extended the search for high-z candidates in Sect.\ \ref{selection-results}
to an adjacent area of $\sim$1000 arcmin$^2$ in the CFHTLS-D3 field, extracted
from the WIRDS Survey in this area \citep{Bielby2012}, and observed only in
$J$, $H$ and $K_s$ bands. This dataset is referenced as ``WIRDS'' in
Table~\ref{tab_phot}, and corresponds to the area delimited in green in
Fig.\ \ref{WUDS_layout}.
 
The seeing in each individual image was determined as the median FWHM of four reference stars. 
The mean and standard deviation values measured across the sample presented in Table~\ref{tab_wuds}
for the different filters are :
0.66\arcsec $\pm$ 0.09\arcsec ($Y$),
0.58\arcsec $\pm$ 0.07\arcsec ($J$),
0.57\arcsec $\pm$ 0.16\arcsec ($H$), and 
0.54\arcsec $\pm$ 0.06\arcsec ($K_s$).
The seeing distribution in $Y$ and $H$ is a little wider than in the two other filters,
leading to a larger {\it rms}. In fact, more than $\sim$90\% of the sample has a 
FWHM better than 0.75\arcsec in these filters, leading to a mean and standard deviation of
0.65\arcsec $\pm$ 0.08\arcsec and 0.53\arcsec $\pm$ 0.06\arcsec in $Y$ and $H$ respectively
when excluding these extreme values. 
The seeing of individual images was included in the weighting process when building
the final stacks, as explained in Sect.\ \ref{reduction}.
Depending on the filter, the averaged seeing values achieved on the stacked
images typically range between 0.55 and 0.63\arcsec, as seen in Table~\ref{tab_phot}. 

   Table~\ref{tab_phot} summarizes the properties of the photometric dataset
used in this paper, when combining the whole WUDS, WIRDS and CFHTLS-D3
observations in this field. The total effective area covered by this survey
with $>$50\% and $>$75\% of the total exposure time is also given in
Tab.\ \ref{tab_phot} for the different bands. The maximum intersection with
all nine filter-bands and $>$50\% of exposure time is limited by the Y-band
(i.e., $\sim$390 arcmin$^2$), whereas it is $\sim$440 arcmin$^2$ with eight 
filter-bands in the extended (WIRDS) area (outside the WUDS region, that is $\sim$830
arcmin$^2$ in total on the CFHTLS-D3 field of view covered at least by 
$JHK_s$ bands with $>$50\% of the total exposure time).  

\begin{table*}
\caption{\label{tab_phot}Photometric dataset used in this paper}
\centering
\begin{tabular}{llrrrccccc}
\hline
Reference & Filter &  $\lambda_{eff}$& $\Delta \lambda_{eff}$ & $C_{AB}$ &
$t_{exp}$ & m(3$\sigma$)& m(50\%) & seeing & Area ($>$50\%) \\
          &        & [nm]           &  [\AA] & [mag]    & [ksec]   & [mag] &
[mag]   & [\arcsec]  & [arcmin$^2$] \\
\hline
CFHTLS-D3 & $u^{\ast}$           &  382 & 544 & 0.312 &  76.6 &  28.52 & 26.97
& 0.89 & 3224 \\
CFHTLS-D3 & $g$           &  490 & 1309&-0.058 &  79.6 &  28.94 & 26.79 & 0.84 & 3224 \\
CFHTLS-D3 & $r$           &  625 & 1086& 0.176 & 142.8 &  28.57 & 26.30 & 0.78 & 3224 \\
CFHTLS-D3 & $i$           &  766 & 1330& 0.404 & 249.4 &  28.24 & 25.95 & 0.76 & 3224 \\
CFHTLS-D3 & $z$           &  884 & 1033& 0.525 & 175.4 &  27.09 & 25.46 & 0.69 & 3224 \\
WUDS      & $Y$           & 1027 & 1077& 0.632 &  44.9 &  26.78 & 26.26 & 0.63
& 392 \\
WUDS      & $J$           & 1256 & 1531& 0.949 &  50.4 &  26.69 & 26.17 & 0.60
& 396 \\
WIRDS     & $J$           &      &     &      &  26.3 &  25.80 &  24.80 &0.60 
& 437 \\
WUDS      & $H$           & 1636 & 2734& 1.390 &  39.6 &  26.06 & 25.61 & 0.55
& 477 \\
WIRDS     & $H$           &      &     &      &  15.5 &  25.73 &  24.80 & 0.55 
& 681 \\
WUDS      & $K_s$         & 2154 & 3071& 1.862 &  25.9 &  25.93 & 25.46 & 0.56
& 450 \\
WIRDS     & $K_s$         &      &     &      &  17.5 &  25.59 &  24.63 & 0.56  
& 547 \\
\hline
\end{tabular}
\tablefoot{Information given in this table:  reference field, filter identification,
filter effective wavelength, filter width, 
AB correction ($m_{AB}=m_{Vega}+C_{AB}$), total exposure time, 3 $\sigma$
limiting magnitudes (within 1.3\arcsec diameter aperture), 50\% completeness
level for point-like sources (in the regions with $>$50\% of the total
exposure time), average seeing for the final stack, and total area covered with
$>$50\% total exposure time. 
}
\end{table*}

\begin{table}
\caption{\label{tab_wuds}WUDS observations summary table}
\centering
\begin{tabular}{lccc}
\hline
Filter & Number     & $t_{exp}$ & Total exposure \\
       &  of cubes  &  [sec]   &     [h] \\
\hline
$Y$ (1) &  561            &  80      &   12.5  \\
$J$ (1) &  400            &  60      &   6.7   \\
$J$ (2) &  589            &  45      &   7.3   \\
$H$ (1) & 1616            &  15      &   6.7   \\
$H$ (2) & 1017            &  15      &   4.3   \\
$K_s$ (1) &  320          &  25      &   2.2   \\
$K_s$ (2) &  874          &  20      &   5.0   \\
\hline
\end{tabular}
\tablefoot{\\
(1) Data from May-July 2008 observations \\
(2) Data from the WIRDS survey on the same area
}
\end{table}

%

\section{Data processing}
\label{reduction}

   Data processing was performed at CFHT (preprocessing phase) and
Terapix/IAP (advanced processing). A two-step approach was adopted for
sky-subtraction and image stacking mostly inspired from
the reduction of near-IR observations by \cite{Labbe03} and \cite{Richard06}
with similar goals. The main steps are the following:

\begin{enumerate}

\item Detrending process of raw images, performed at the CFHT. This includes flagging the
  saturated pixels, correcting for non-linearity, reference pixels
  subtraction, dark subtraction, dome flat-fielding, bad-pixels masking, and
  guide-window masking. These steps are described in details at the WIRCam
  home page at CFHT. 

\item First sky-subtraction. Given the fast variations of the sky-background
  on large and small spatial scales, we used all the science images taken between $\sim$10
  minutes before and after the actual image to produce a ``sky''
  background, by medianing these adjacent exposures. The efficiency of this
  process strongly depends on the dithering strategy, that is using larger
  offset-paths provides better results. 

\item Astrometry and photometric calibration (standard preprocessing; see below).

\item First image stack. Images were sky-subtracted using the previous ``local''
  backgrounds and then registered and combined together into a first stack.

\item Object mask. Sources were detected with SExtractor \citep{Sex} in the
  first image stacks in order to create an object mask. 

\item Second sky-subtraction. The second step was repeated using the object
  mask to reject pixels located on detected sources when computing the sky value.
  A second sky-subtraction was applied to the data using these new backgrounds. 

\item Final stack. Before combining the frames into a final stack, we applied
  weight values to individual images optimized to improve the detectability of
  faint compact sources in this way: $weight \propto (ZP\times var\times
  {s}^2)^{-1}$, where $ZP$ and $s$ correspond to zero-point and seeing values
  respectively, and $var$ is the pixel-to-pixel variance derived in a
  reference clean area. 

\end{enumerate}

   Automatic preprocessing at CFHT included steps from 1 to 3 above. It was
done with the 'I'iwi IDL Interpretor of WIRCam Images (see details at the
WIRCam home page at CFHT). Astrometry included the detection of stars using
SExtractor, the computation of a full mosaic WCS linear solution, followed by a
detector-by-detector refinement using IMWCS. For each filter-band, photometric
calibration was performed by WCS matching all stars detected by SExtractor to
the 2MASS photometry, on a detector by detector basis. For the $J$, $H$ and
$K_s$ bands, ZPs were derived from reference 2MASS stars in the Vega system
\citep{Skrutskie06}, converted into AB magnitudes using the conversion values
in Table~\ref{tab_phot}. For the $Y$ band, a
ZP was estimated using reference spectrophotometric stars. The first images
obtained in this way suffered from several problems described below, and were
used only for tests as first-epoch data.  

\begin{figure}
   \centering
   \includegraphics[width=0.48\textwidth]{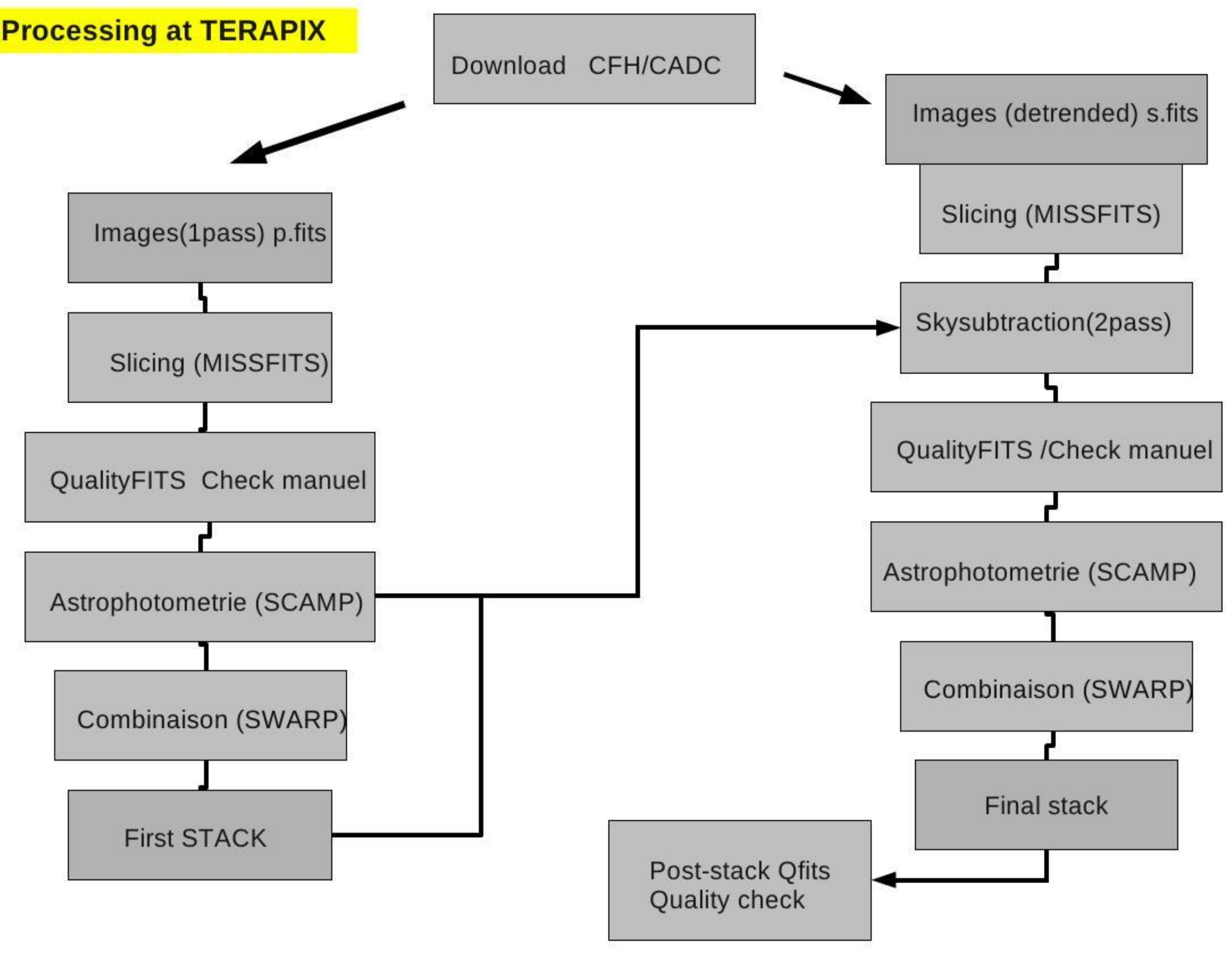}
      \caption{Schematic view of the advanced processing of WUDS data at Terapix.}
         \label{Terapix_proc}
   \end{figure}

   The advanced processing at Terapix started from detrended images, therefore
including steps from 2 to 7 above. Fig.~\ref{Terapix_proc} presents a
schematic view of this process. Two particular problems needed a specific
solution to improve the final stack. Firstly, a large fraction of images suffered from
horizontal stripe-like residuals due to the detector amplifier, with a small
amplitude (typically 10 counts over $\sim$10000 counts for the sky
background). They were successfully removed from the individual images by
subtracting a model obtained from the horizontal projection of thin stripes
($\sim$1/4 of the amplifier width), after object masking. The second
correction was performed to suppress large-scale gradient residuals from the
background sky. This correction was obtained through SExtractor background
subtraction, using a large mesh-size (256 pixels) on images where objects had
been previously masked. It is worth to note the highly time-consuming process of 
manual quality-assessment of individual images (using QualityFITS), given the
huge number of images in the stacks ($\sim$560 in $Y$, $\sim$1420 in $J$,
$\sim$3670 in $H$ and $\sim$1920 in $K_s$), and the fact that the whole
process was performed twice. Also weight maps were obtained during this phase, using the
WeightWatcher software developed by \cite{2008ASPC..394..619M}. 

   Astrometric calibration was performed with 
SCAMP\footnote{\tt{http://www.astromatic.net/software/}}. 
The accuracy of internal astrometry ranges between 0.02 and 0.035\arcsec at
1$\sigma$ level for all filters. Images were combined
using the weighting scheme given in point 7 above, with the SWARP
software. The combination was a sigma-clipped mean with a 3$\sigma$ rejection
threshold. The final stacks were matched to the CFHTLS-D3 T0006 images and pixel
scale (0.186\arcsec). The astrometric solution was computed by SCAMP 
including internal constraints (overlapping frames) and external references
(CFHTLS-D3 objects catalog). The accuracy of astrometry in this case is typically
$\sigma$=0.05\arcsec, with a systematic offset of less than 0.005\arcsec. 

   Regarding the accuracy achieved in the determination of ZPs, it ranges from 0.021
to 0.023 magnitudes in $J$, $H$ and $K_s$ bands in the WUDS survey
alone (it is between 0.03 and 0.04 when combining WUDS and WIRDS data on the
whole field). It is 0.067 magnitudes in the $Y$-band for WUDS (0.075 for WUDS
$+$ WIRDS) due to indirect recalibration. These estimates are
based the comparison between individual detections of several 10$^4$ stars in
each band. 

%

\section{WUDS Multiband Catalogs}
\label{catalog_built}

This Section presents the extraction of sources and the construction of the photometric catalogs,
publicly available at \url{http://wuds.irap.omp.eu/} 

To built the final catalogs, WUDS data were combined with public
CFHTLS-Deep data on this field, obtained through five bands in the
optical domain, namely $u^{\ast}$, $g$, $r$, $i$, and $z$.
Therefore, WUDS catalogs include nine filter bands with full wavelength
coverage between $\sim$0.35 and 2.3 $\mu$m.
Hereafter, we refer to this 
ensemble as WUDS (or WUDS$+$WIRDS) data. 


   Two near-IR-selected catalogs were built for the needs of this
project. The first one (hereafter C1) is based on the $Y+J$ detection image. 
It was primarily intended to be used for the identification of $i$ and $z$
dropouts in this paper (i.e., z$\sim$6 and z$\sim$7
candidates respectively). The second one (C2) is based on the $H+K_s$
detection image, as it was intended to be used for the identification of $Y$
and $J$ dropouts (i.e., z$\sim$7-11 candidates).
This later selection will be presented in Paper II. 
We also computed a catalog based on the $i+z$ detection image to be used for
the identification of $r$-dropouts (i.e., z$\sim$5 candidates) in addition to
the near-IR selected samples. 
All these catalogs are used in Sect.\ \ref{selection-results}
for the selection of galaxies at z$\sim [$4.5 -- 7$]$. 

   Sources were detected with the SExtractor package version 2.8
\citep{Sex}, using the weight maps mentioned in Sect.~\ref{reduction}.
Extraction was performed using a very low detection threshold of
0.8 sigma (SExtractor definition) for a minimum number of four pixels
above the threshold, in order to optimize the detection of compact and
faint sources.
A background mesh of 64 pixels was used for background
subtraction. Magnitudes and fluxes were measured in all images with the
``double-image'' using the corresponding detection 
images ($i+z$, $Y+J$ or $H+K_s$). Total magnitudes and fluxes were computed
based on SExtractor MAG\_AUTO magnitudes. Also aperture magnitudes were
derived within 14 different apertures ranging from 1.3 to 5\arcsec diameter,
on the original images, and also on images matched to the $u^{\ast}$-band
seeing using a simple Gaussian convolution.  
Photometric errors were measured using the typical background
variance of the original images (without any seeing matching or rescaling),
within apertures of the same physical size as for flux measurements (either
aperture or MAG\_AUTO magnitudes). Errors in colors were derived by
quadratically adding the corresponding errors in magnitude.

\begin{figure*}
   \centering
   \includegraphics[width=0.48\textwidth]{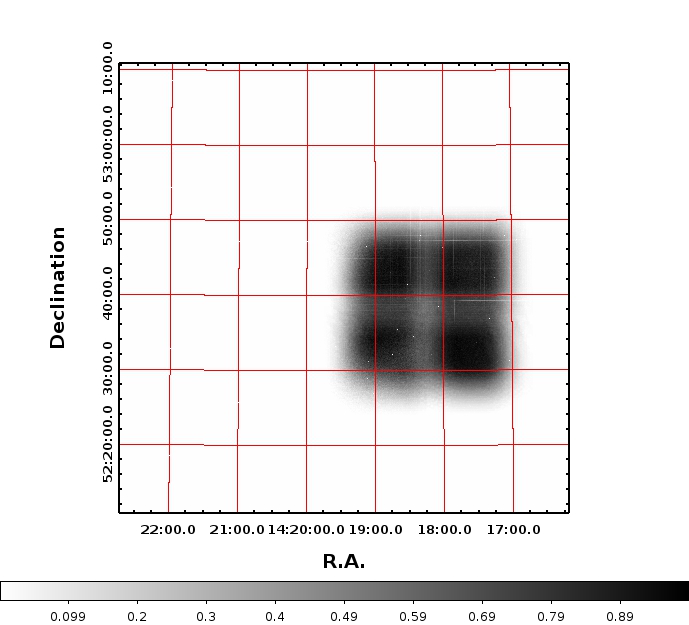}
   \includegraphics[width=0.48\textwidth]{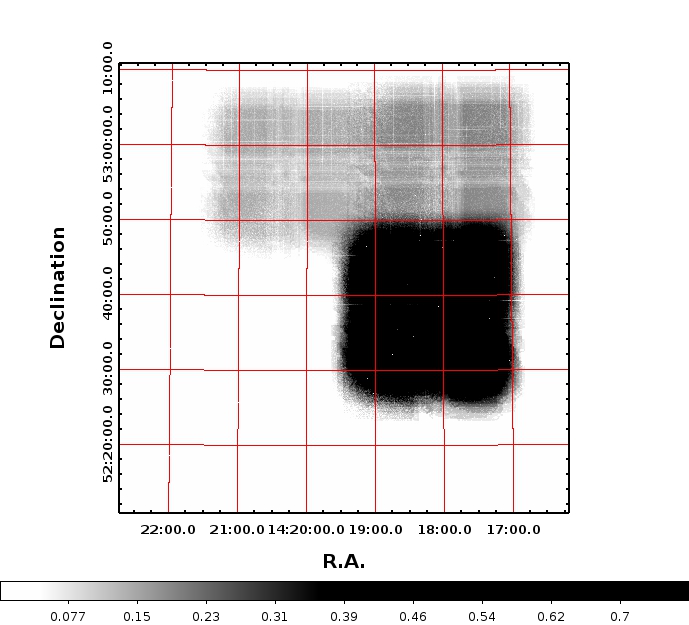} \\
   \includegraphics[width=0.48\textwidth]{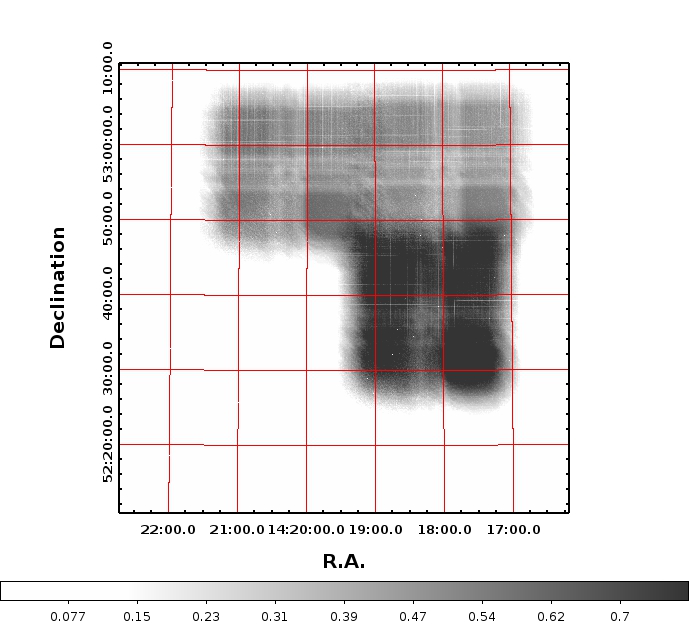}
   \includegraphics[width=0.48\textwidth]{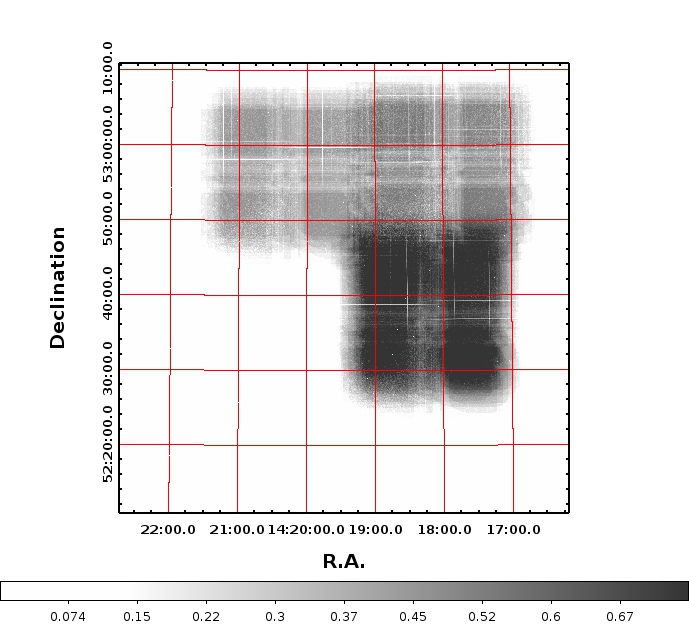} 
   \caption{Exposure-time maps for the CFHTLS-D3 field of view in $Y$ (top left),
        $J$ (top right), $H$ (bottom left) and $K_s$ bands (bottom
        right). Gray levels display in linear scale the regions where near-IR data are
        available. Note the different exposure times between the WUDS (deep)
        region and the wide field extended region of WIRDS.   
      }
         \label{WUDS_weight}
   \end{figure*}

Figure~\ref{WUDS_weight} displays the exposure-time maps 
for the $\sim$1$\times$1\degr \ CFHTLS-D3 field of view, and for the different near-IR 
filters. We note the difference in the total exposure times between the WUDS
(deep) region and the extended region of WIRDS. Table ~\ref{tab_detections}
summarizes the number of sources detected in the different areas, for the
different filters and detection images. Table~\ref{tab_phot} also presents
the 3$\sigma$ limiting magnitudes achieved in the different filters within
1.3\arcsec diameter aperture (point sources). 
Completeness and depth achieved by WUDS are discussed below
(see Sect.~\ref{completeness}).  

\section{Characterization of the photometric survey}
\label{catalogs}

In this Section we characterize the properties of the WUDS photometric catalogs in
different ways. Completeness and
depth are estimated based on realistic simulations of stars and
galaxies. Number counts are obtained in the near-IR bands and compared to
previous findings. 

\begin{table*}
\caption{\label{tab_detections} Summary of detections in the main survey
 (WUDS) and extended area (WIRDS), for the different near-IR filters and
 detection images (1).
Detection images reported in column (1) are the same for a given row.
A column is provided for each filter reporting the total   
number of sources brighter than the magnitude given in the second row
(see discussion in Sect.\ \ref{completeness}).
The last two columns refer to objects simultaneously detected in the three filters $JHK_s$ and
the four filters $YJHK_s$ respectively.    
}
\centering
\begin{tabular}{lccccccc}
\hline
Area & & $Y $        & $J $        & $H $        & $ K_s$      & $JHK_s$ & $YJHK_s$  \\
     &(1) & AB $< 25.5$ & AB$ < 25.5$ & AB$ < 24.75$ &AB $< 24.75$&         & \\
\hline
WUDS & $Y+J$&    77715 & 91340 & 59552 & 66623 & 34832 & 29841 \\
     & $H+K_s$&  66479 & 79526 & 66627 & 72960 & 35791 & 30655 \\
     & $i+z$&    49920 & 57574 & 46185 & 49089 & 31900 & 28088 \\
WIRDS & $Y+J$&     -   & 82200 & 35999 & 46225 & 21721 &   -   \\
     & $H+K_s$&    -   & 86985 & 48063 & 65858 & 21706 &   -   \\
     & $i+z$&      -   & 41637 & 31759 & 36056 & 18371 &   -   \\
\hline
\end{tabular}
\end{table*}

%

\subsection{Completeness and depth}
\label{completeness}

   The completeness of the WUDS survey has been estimated through 
simulations of stars and galaxies based on the STUFF and SKYMAKER
softwares. Simulated samples of stars/galaxies have been randomly added to
real images after masking of objects detected in the stacks, with the
appropriate PSF convolution. These sources are then detected with SExtractor,
using the same extraction parameters as for science images, therefore the
completeness levels are directly obtained as a function of magnitude in the
different bands. These values are reported in Tab.\ \ref{tab_phot}.
The difference between the completeness levels in regions with the
highest exposure time ($>$90\% ) and regions with smaller exposure
times, for instance regions with $>$70\% and  $>$50\% exposure time, is typically 0.08
and 0.13 magnitudes worse respectively. Therefore, given the geometry of the survey,
the search for optical dropouts is limited in practice to regions with at
least 50\% of the total exposure time in the final stack (see Tab.\ \ref{tab_phot}).
This means that the detection level does not
change dramatically across the surveyed field. 

   The typical depth of the WUDS data reaches between $\sim$26.8 in $Y$ and $J$, 
and $\sim$26 in $H$ and K$_s$ (AB, 3$\sigma$ in 1.3\arcsec 
aperture), for a completeness level of $\sim$80\% at $Y\sim$26 and $H$ and
K$_s$$\sim$25.2, and excellent seeing values for the final stacks (ranging
between 0.55 and 0.65 arcsec). The corresponding depth of the CFHTLS-D3 images
in this region ranges between 28.6 and 29 in $ugr$, 28.2 in $i$ and 27.1 in
$z$ (same S/N ratio and aperture). 

  The price to pay for the low detection threshold used in this survey is an
enhanced fraction of spurious detections, increasing with magnitude. As
explained in Sect.\ \ref{counts} below, we have estimated this contamination using
the same detection scheme on ``negative'' images and, in addition,
all high-z candidates have been manually inspected. Based on these results, we
have estimated that our catalogs are robust (in the sense that they are not dominated
by false-positive detections in the near-IR images) up to AB$\sim$ 25.5 in $Y$
and $J$, and AB$\sim$24.75 in $H$ and $K_s$, irrespective of the detection image. 
The number of sources detected in the different areas, with magnitudes brighter
than these limits are reported in Table ~\ref{tab_detections}, for the
different filters and detection images.

\subsection{Number counts}
\label{counts}

    Galaxy number counts have been obtained in the four bands of WUDS and
compared to the literature, as a consistency check for the calibration of the present
data set. The separation between stars and galaxies is based on the SExtractor
stellarity 
index\footnote{This index ranges between 0.0 for extended sources and 1.0 for
unresolved ones \citep[see][]{Sex}.}.
Since the reliability of this index for galaxies diminishes toward the faintest
magnitudes, we have applied this selection up to a S/N$\sim$10 in the detection
images, with galaxies selected by imposing a SExtractor stellarity index <0.9.
These threshold values are based on straightforward simulations, using the same
approach as for the determination of the completeness levels in Sect.\ \ref{completeness}.

\begin{figure*}
   \centering
   \includegraphics[width=0.24\textwidth]{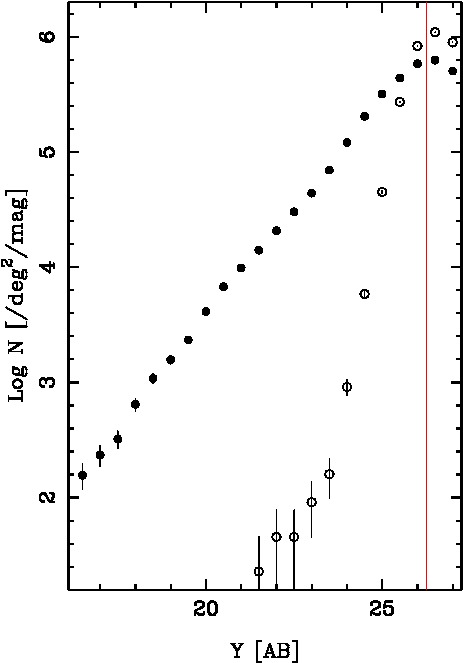}
   \includegraphics[width=0.24\textwidth]{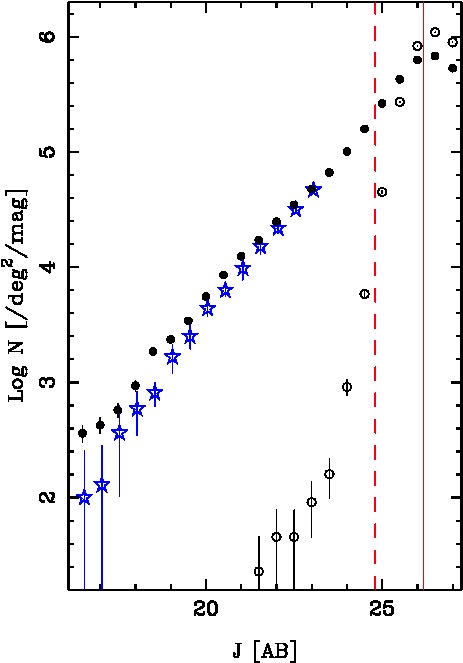} 
   \includegraphics[width=0.24\textwidth]{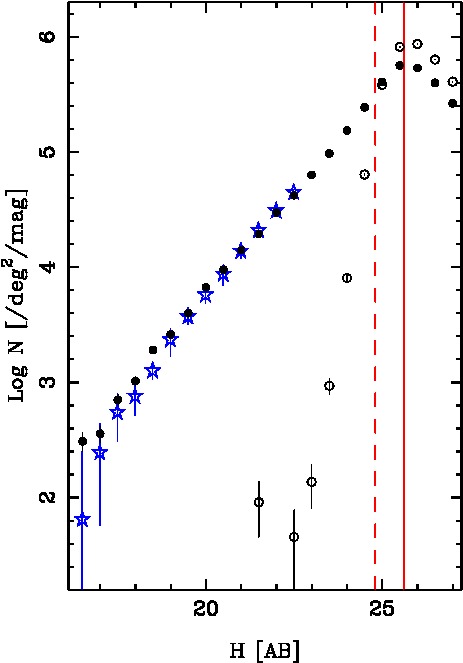}
   \includegraphics[width=0.24\textwidth]{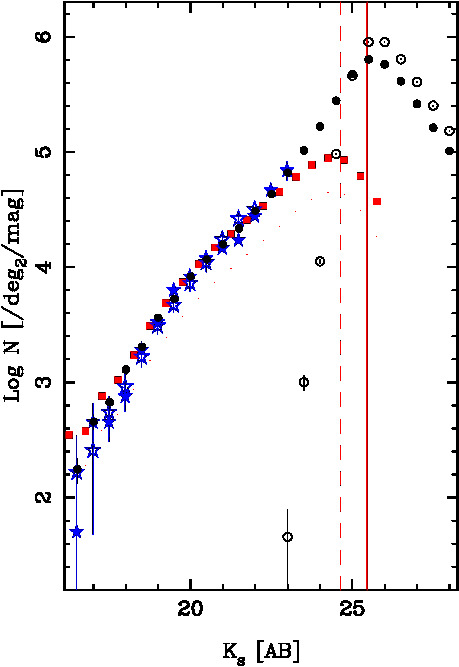} 
      \caption{Galaxy number counts in the four bands of WUDS. Black filled
circles display the results in the WUDS field. Results from \cite{Cristobal09}
(open stars), \cite{Cristobal03} (filled stars) and
\cite{2016ApJS..224...24L} (red points) 
are also shown for comparison. Open circles display the results
obtained on the negative images, as an indication of contamination
by spurious sources. Error bars correspond to 1$\sigma$ Poissonian errors.
Vertical solid and dashed lines indicate the 50\% completeness levels
for point sources in WUDS and extended (WIRDS) areas respectively. 
      }
         \label{WUDS_counts}
   \end{figure*}

     Figure \ref{WUDS_counts} displays the resulting number counts in the four 
bands of WUDS, as compared to previous findings from the literature
\citep[e.g.,][]{Cristobal03,Cristobal09,2016ApJS..224...24L}, without any correction for incompleteness. 
The detection image was $Y+J$ for $Y$ and $J$ bands, whereas it was $H+K_s$
for $H$ and $K_s$. The 50\% completeness levels for point sources coincide
with the drop in number counts in the WUDS area. 
Compact sources with magnitudes brighter
than 17.7 in $Y$, 18.1 in $J$, 17.2 in $H$ and 17.6 in $K_s$ are affected by
saturation. Although there is some scatter
toward the bright end (AB\ltapprox 20), there is a good agreement with
previously published results, in particular \cite{Bielby2012}. 

    The detection scheme described above was optimized to identify faint and compact sources
(see Sect.\ \ref{completeness}). Therefore, a large fraction of spurious
detections was expected in the near-IR bands, increasing toward the faintest
magnitudes. In order to evaluate the incidence of this effect, we
have applied an identical scheme for source detection as described in Sect.\ \ref{completeness} 
to negative images obtained by multiplying the original stacks by $-1$, to
blindly extract these spurious non-astronomical signal. The result of this procedure is also shown in
Fig.~\ref{WUDS_counts} (open dots). The structure of the noise in these
negative images is somewhat different with respect to the astronomical
ones, in the sense that an excess of faint and compact sources appears toward the faintest
magnitudes. As seen in Fig.~\ref{WUDS_counts}, this systematic trend and the
dominance of false positives start close to the 50\% completeness levels in WUDS. 
The reason for this trend, which is also observed in other similar
surveys (e.g., public CLASH data from HST), is not clear. It could be due to
the drizzling and resampling procedure. For this reason, we did not try to use
these negative counts to correct our results, but as an indication of the
flux level at which severe contamination is expected. 
In practice, we have limited the detection samples to magnitudes reported in
Table ~\ref{tab_detections}, where contamination is
not expected to dominate. In addition, all the high-z candidates presented in this paper and in
Paper II have been manually inspected to remove obvious spurious sources.

    Regarding the comparison with previous findings on near-IR counts, our results
are fully consistent with \cite{Cristobal03}, \cite{Cristobal09} and
\cite{2016ApJS..224...24L} 
at m$>$20, whereas a larger dispersion is observed for brighter sources as expected.
WUDS is still deeper than UltraVISTA-DR2 survey in the $K_s$-band \citep{2016ApJS..224...24L}. 
The change in the slope of the near-IR counts at AB$\sim$19.5-20.0 is clearly visible
in the $K_s$-band, whereas it is less obvious in the other bands, and not present in the
optical bands \citep[e.g.,][]{Cristobal09,2006ApJ...639..644E}. 
This trend has been identified by several authors using models   
\citep[e.g.,][and the references therein]{2010A&A...519A..55E, 
2015MNRAS.451.1158P} 
as the result of the late assembly by major-mergers of a substantial fraction
of present-day massive early-type galaxies at 0.7$<$z$<$1.2, and inconsistent
with a simple passive evolution since z$\sim$2. 

%

\section{Data properties at low-$z$}
\label{data_lowz}

In this Section we assess the quality achieved in the computation of
photometric redshifts based on these data, up to z$\sim$1.5. We also 
use the SED-fitting approach to
derive the properties of the near-IR selected galaxy population in this
field in terms of redshift distribution and stellar masses.

\subsection{Photometric redshifts}
\label{photoz}

   Photometric redshifts have been computed with the version v12 of the public code
{\it Hyperz\/} 
({\it New$-$Hyperz\/}\footnote{http://userpages.irap.omp.eu/$\sim$rpello/newhyperz/}), 
originally developped by \cite{hyperz}. This method is based on the fitting of
the photometric Spectral Energy Distributions (SED) of galaxies. The accuracy of
photometric redshifts (\zphot) is used here as a consistency check for the
calibration of the whole data set, as well as for the characterization of the
different high-z samples in WUDS. 

   The template library used in this paper includes 14 templates: 
eight evolutionary synthetic SEDs computed with the last
version of the Bruzual \& Charlot code \citep{BC2003}, 
with Chabrier IMF \citep{Chabrier2003} and solar metallicity, 
matching the observed colors of local galaxies from E to Im types 
(namely a delta burst -SSP-, a constant star-forming
system, and six $\tau$-models with exponentially decaying SFR);
a set of four empirical SEDs compiled by \cite{CWW},
and two starburst galaxies from the \cite{Kinney1996} library.
Internal extinction is considered as a free parameter following the
\cite{Calzetti2000} extinction law, with A$_V$
ranging between 0 and 3.0 magnitudes (E(B-V)
in the range $\sim$[0,0.75] mag).

   Photometric redshifts have been computed in the range z$=$[0,12] using two
different priors in luminosity. The first one is the usual flat luminosity
prior in redshift, that is a simple cut in the permitted range of luminosities
for extragalactic sources, with absolute magnitudes in the range
M$_B$=$[$-14,-23$]$. The second one is a ``soft'' probability
distribution as a function of redshift and magnitude, following
\cite{Benitez2000}, encompassing the $B$-band luminosity function derived by
\cite{Ilbert2006}. This new option of {\it New$-$Hyperz\/} computes 
a smooth probability distribution prior for each object as a function of
redshift, the absolute magnitude M$_B$ being derived from the apparent
magnitude $m$ which is closer to the rest-frame $B$-band. The final
probability distribution is given by the usual $Hyperz$ $P(z)$ combined with
the prior. 

   {\it New$-$Hyperz\/} performs a $\chi^2$ minimization in the
parameter space in the first pass, followed by a small correction for
systematics trends as a function of \zphot obtained through the polynomial fit
of the residuals between the best-fit redshift above and the true value for the
spectroscopic sample described below, excluding outliers. These residuals
encode our lack of precise knowledge on the overall system transmission as a
function of wavelength. 
The procedure yields the best fit \zphot and model template for each
source, as well as a number of fitting byproducts (e.g., absolute magnitudes
in the different bands, normalized redshift probability distribution, \zphot
error bars, secondary solutions, ...). An interesting indicator of the
goodness of the fit is provided by the integrated probability \Pint between
\zphot$\pm$0.1, where \zphot stands for the best fit redshift, with the
probability distribution normalized between z$=$[0, 12]. 
Among the fitting byproducts is a rough classification of the rest-frame SED
of galaxies into five different spectral types, according to their best fit with
the simplest empirical templates given by \cite{CWW} and \cite{Kinney1996}:
(1) E/S0, (2) Sbc, (3) Scd, (4) Im and (5) S (starbursts).

\begin{figure}
   \centering
   \includegraphics[angle=0,width=0.48\textwidth]{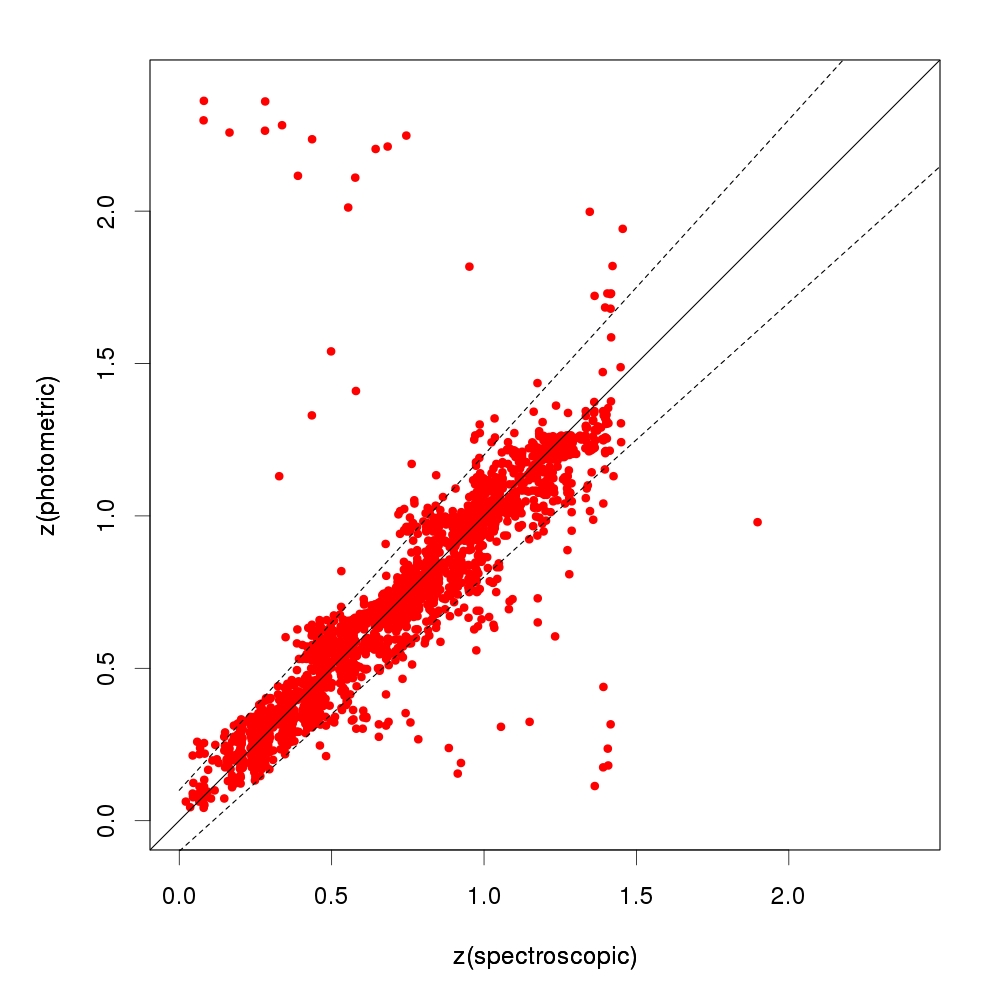}
   \includegraphics[angle=0,width=0.48\textwidth]{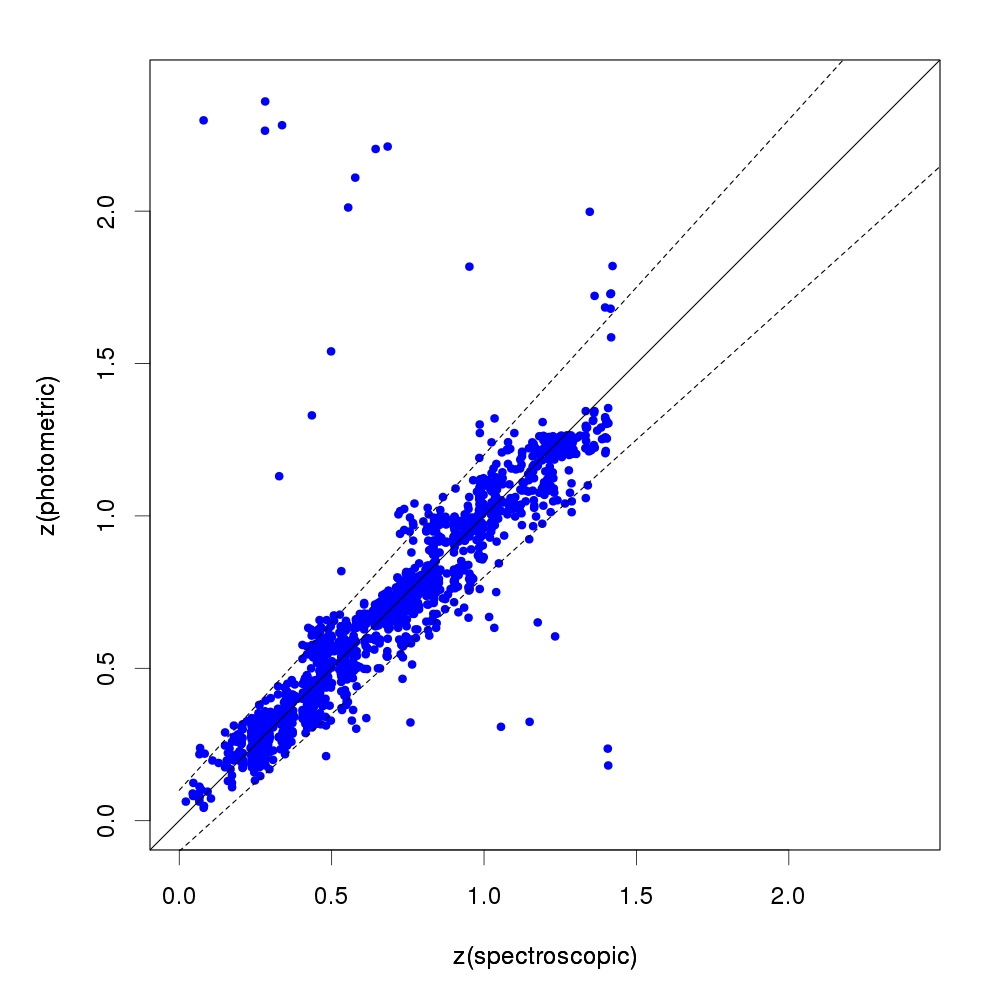}
      \caption{Comparison between photometric and spectroscopic redshifts in
this survey for the $H+K$ detection image across the entire field (top), 
and for the WUDS area only (bottom). 
Dot-dashed lines display the locus of $z(phot)=z(spec)\pm 0.1(1+z)$ to guide
the eye.  
      }
         \label{fig_zz}
   \end{figure}

    The photometric redshift accuracy has been estimated through a direct
comparison between \zphot and secure spectroscopic samples publically
available in the CFHTLS-D3 field, extracted from the DEEP Groth Streep Survey DR3
\citep{Weiner2005,Davis2003,Davis2007}. Photometric and spectroscopic catalogs
were blindly matched in ALPHA and DEC positions. Only objects with the highest
spectroscopic redshift quality (ZQUALITY $\geq$ 3) were considered in this
comparison, that is 3424(3409) galaxies in the entire WUDS$+$WIRDS area
based on $H+K$($Y+J$) detection images. Magnitudes in this sample range between
$i$=18 and 24.4, with median value $i$=22.4, corresponding to $H$=$\sim[$17.0 -- 24.8 $]$
with median value $H$=21.9. Results based on $H+K$ detection are
the same as for the $Y+J$-based catalog. We have also considered the 3828 galaxies
extracted form the $i+z$ detection image. Figure ~\ref{fig_zz} displays the
comparison between photometric and spectroscopic redshifts based on the
$H+K$ detection image across the entire field, and the same restricted to
the WUDS area, that is with photometry including $Y$-band data and 1651
spectroscopic sources.

    Table ~\ref{tab_zz} presents a summary of the \zphot quality achieved in
this survey based on the usual statistical indicators, namely $\sigma(\Delta
z/(1+z))$, $\sigma(|\Delta z/(1+z)|)$, the median of $(\Delta z/(1+z))$, 
the normalized median absolute deviation (defined as
$\sigma_{z,MAD}$ = 1.48 $\times$ median $(|\Delta z|/(1+z))$, which is less
sensitive to outliers), and the percentage of outliers. 
These results are based on SExtractor MAG\_AUTO
magnitudes. Outliers are defined here as sources with $|z(spec)-z(phot)| >
0.15(1+z(spec))$. As shown in the table, the dispersion is below or on the order of
$\sim 0.05(1+z)$ in all cases based on the usual indicators, and the
percentage of outliers ranges between 4 and 5\% for the entire field,
improving to 3-4\% for the WUDS area, the results being slightly
better for the near-IR selected samples. The same trends are also seen in
Fig.~\ref{fig_zz}, although the difference is small and hardly
significant. The correction for systematic trends mentioned above is included
in the results presented in Table ~\ref{tab_zz} and Fig.~\ref{fig_zz}. When
this correction is not included, the results on the dispersion are worse by
$\sim0.002$ to 0.006 depending on the sample and indicator, whereas the median
bias $(\Delta z/(1+z))$ is up to a factor of 10 larger. 

\begin{table}
\caption{\label{tab_zz} Summary of the \zphot quality achieved with {\it
    New$-$Hyperz\/} on the WUDS/CFHTLS D3 field.}
\centering
\begin{tabular}{lcccc}
\hline
Detection & $H+K$ & (all)  &  $i+z$ & (all) \\
          &  (1)  & (2) &  (1)   & (2)  \\
\hline
$\sigma(\Delta z/(1+z))$ & 0.047 & 0.048 & 0.051 & 0.050   \\
$\sigma(|\Delta z/(1+z)|)$& 0.031 &0.031 & 0.032 &  0.032  \\
Median (3) & 0.0016 & 0.0017 & -0.0017 & -0.0012 \\
$\sigma_{z,MAD}$ & 0.043 & 0.043 & 0.046 & 0.045 \\
Outliers & 4.4\% & 4.4\% &  4.9\% & 4.9\%  \\
\hline
Detection & $H+K$ & (WUDS)    &  $i+z$ & (WUDS) \\
          &  (1)  & (2) &  (1)   & (2)  \\
\hline
$\sigma(\Delta z/(1+z))$ & 0.047 & 0.047 & 0.049 & 0.049   \\
$\sigma(|\Delta z/(1+z)|)$& 0.030 &0.030 & 0.031 &  0.031  \\
Median (3) & 0.0005 & 0.0007 & 0.0007 & 0.0007 \\
$\sigma_{z,MAD}$ & 0.042 & 0.042 & 0.044 & 0.045 \\
Outliers & 2.9\% & 2.8\% &  3.8\% & 3.9\%  \\
\hline
\end{tabular}
\tablefoot{\\
(1) Flat prior \\
(2) LF prior \\
(3) Median $(\Delta z/(1+z))$
}
\end{table}

   The availability of near-IR filters helps improving the \zphot accuracy
beyond $z\sim$1.3, where the 4000\AA \ break goes out of the $z'$ filter and
the Lyman break is not yet detectable in the $u^{\ast}$ band. The main impact
when including $Y$-band data is on the percentage of outliers. 
Unfortunately, only $\sim$3\% of the spectroscopic control sample is found at $z\ge$ 1.3. 
Results obtained with a flat luminosity prior are not significantly different from
those achieved using a more aggressive prior based on the LF. Results based
on seeing-matched aperture magnitudes taking the $u^{\ast}$-band as a
reference are significantly worse than those based on MAG\_AUTO, with a
dispersion increasing by ~0.01 to 0.02 with respect to the $\sigma$ values
displayed in Table ~\ref{tab_zz}. The reason for this is that the improved sampling in the
SED obtained for seeing-matched apertures is compensated by a worse S/N ratio
in the photometry as compared to MAG\_AUTO, the net effect being a lower
quality in the photometric redshifts.

The final quality achieved for WUDS without further refinement is within
the requirements for large cosmological surveys \citep[e.g., Euclid,][]{Laureijs11},
and it is expected to be representative of the general behavior for other
SED-fitting \zphot codes applied to these data. Indeed, the dispersion and the percentage of
outliers are comparable to the ones obtained with a similar number of filters
in this redshift domain, such as for the 22.5$< i < $24.0 -selected sample in
the VIMOS VLT Deep Survey (VVDS) \citep{2006A&A...457..841I}, 
the Hubble Ultra Deep Field \citep{2006AJ....132..926C}, 
or the COSMOS field \citep{2007ApJS..172..117M}, 
whereas a better accuracy could be achieved by using a wider filter set 
\citep[see e.g.,][]{2009ApJ...690.1236I}. 

\subsection{Properties of the galaxy population}
\label{properties}

   We have taken advantage from the SED-fitting byproducts obtained by {\it
New$-$Hyperz\/} when deriving \zphot, as described in Sect.~\ref{photoz},  
to characterize the properties of the galaxy population in the WUDS survey. 
These quantities have been computed based on the catalogs used to select high-z
sources in Sect.~\ref{selection-results} below. Hereafter in this section, we limit the sample
to sources fainter than the saturation limits in all WUDS filters, 
with SExtractor stellarity index $<$0.9, 
and detected in at least two near-IR filters with magnitudes brighter than
AB$\sim$25.5 in $Y$ and $J$, and AB$\sim$24.75 in $H$ and $K_s$. As discussed
in Sect.~\ref{counts}, these limits ensure that the sample is
not dominated by spurious sources. The sample presented here contains
$\sim$110(118) x 10$^{3}$
sources in both catalogs based on $Y+J$($H+K_s$) detection images. 

\begin{figure}
   \centering
   \includegraphics[angle=0,width=0.48\textwidth]{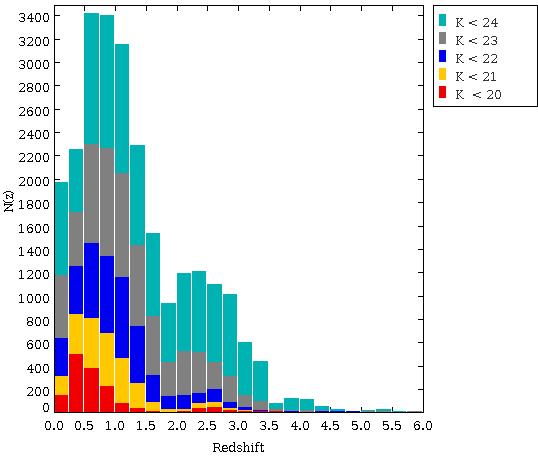}
      \caption{Comparison between the photometric redshift distribution 
obtained for different $K_s$-band selected samples in WUDS, from
top to bottom $K_s < $ 24, 23, 22, 21 and 20. 
      }
         \label{histo_z_KAB}
   \end{figure}

   Figure ~\ref{histo_z_KAB} displays the photometric redshift distribution
obtained for different $K_s$-band selected samples in this survey. As expected,
the distribution extends to higher redshifts with increasing magnitudes, with a
clear drop in the distribution of galaxies at z\gtapprox 3.5, when the
4000\AA \ break enters the $K_s$-band. 

\begin{figure*}
   \centering
   \includegraphics[angle=0,width=0.95\textwidth]{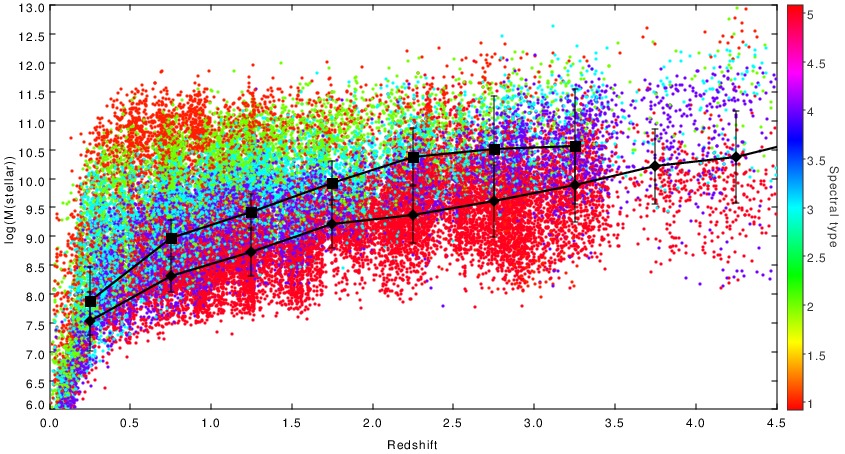}
      \caption{Distribution of stellar masses as a function of redshift for
the catalog based on $H+K$ detection image. Colors encode the different
spectral types of galaxies, from (1) E/S0 to (5) Starbursts.
Black squares and diamonds represent the
completeness limits in mass up to $K_s$ = 24.75, for early and late-type
galaxies respectively. Error bars represent the dispersion within the
sample for a typical $K_s =$ 24.75 galaxy (see text for details). 
      }
         \label{logM_z}
   \end{figure*}

   Stellar masses are among the quantities derived by the SED-fitting
procedure, based on the best-fit model obtained with the
Bruzual \& Charlot code \citep{BC2003}, with Chabrier IMF \citep{Chabrier2003}
and solar metallicity. The parameter space is precisely the same used for
\zphot. As discussed by \cite{Davidzon2013}, stellar masses derived
in this way depend very weakly on the parameter space used for SED-fitting, in
particular the detailed star-formation histories.
A rough classification for the rest-frame SED of galaxies 
is provided by {\it New$-$Hyperz\/} using five arbitrary spectral types,
according to their best fit with the simplest empirical templates in the local
universe, namely (1) E/S0, (2) Sbc, (3) Scd, (4) Im and (5) Starbursts
(see Sect.~\ref{photoz}). 
We consider here as genuine ``early-type'' all galaxies with best-fit type
(1), and all the others are considered as ``late-type''. Following
\cite{Pozzetti2010}, we have determined the completeness in mass for the early and
late-type galaxies respectively for a sample limited to $K_s\le$24.75. This was done by
computing the mass it would have an early/late type galaxy at the center of each
redshift bin if its apparent magnitude was $K_s$=24.75.
Figure ~\ref{logM_z} displays the distribution of stellar masses measured in the
WUDS field as a function of redshift and galaxy types, together with the limiting
mass corresponding to $K_s =$ 24.75 for early and late-type
galaxies. Error bars in this figure represent the dispersion
within the sample for a typical $K_s =$ 24.75 galaxy.

Several trends in Figure ~\ref{logM_z} deserve a specific comment.
There is a systematic trend in the sense that the dispersion within the sample increases with
redshift for a given spectral type.
Uncertainties in the determination of individual stellar masses due to various degeneracies
in the parameter space are expected to be typically below 0.2 dex up to z$\sim$3.5,
that is when the observed SED includes the 4000\AA \ break, but the determination becomes
hazardous beyond this redshift 
\citep[see e.g.,][for a detailed discussion]{2013MNRAS.435.1389P, 
2013MNRAS.435...87M}. 
As seen in Figures ~\ref{histo_z_KAB} and ~\ref{logM_z}, the population of galaxies
beyond z$\sim$3.5 strongly diminishes in our near-IR selected catalog, essentially
because the region of the SED beyond the 4000\AA \ break, tracing the  
``old'' stellar population and the stellar mass, progressively moves beyond the reddest
band ($K_s$). This trend is expected given the limited wavelength coverage of 
WUDS (the reddest band is $K_s$). 
For this reason, the completeness limit displayed in Figure ~\ref{logM_z} for
early-type galaxies stops at z$=$3.5, whereas it is given as an indication only for late-type
galaxies beyond this limit, given the uncertainties associated to the stellar mass
determination beyond this redshift. 

   Although WUDS catalogs were built to fulfill the needs of our primary
science goal around the z$\ge$4.5 population, they could also be advantageously used for
studies at low and mid-z, in particular for the stellar-mass preselection of galaxies for
spectroscopic studies.
Up to z$\sim$3.5, early-type galaxies can be reliably identified and used for statistical
purposes down to log(M$^*$/M$\odot$)$>$10.5$\pm$1.0 (completeness for a typical $K_s$=24.75 galaxy),
whereas we expect to detect late-type galaxies down to log(M$^*$/M$\odot$)$>$9.9$\pm$0.60.
Beyond this limit in redshift, stellar masses cannot be properly determined. 

\section{Galaxies at z$\sim [$4.5 -- 7$]$}
\label{selection-results}
\subsection{Selection Criteria}
\label{selection}

  We have used two different methods to select high-z galaxies. 
The first one is the usual three-band dropout technique applied to a
combination of WUDS data and deep optical data from the CFHTLS-D3,
that is a photometric catalog including the nine filter-bands ($ugrizYJHK_s$).
Different redshift intervals have been defined using an appropriate combination
of filters. The second method is based on pure photometric
redshifts and probability distributions, taking full advantage from the wide
wavelength coverage. 
In all cases, a S/N higher than 5$\sigma$ was requested in the filters
encompassing the rest-frame UV, irrespective of the detection image, together
with a non-detection ($<$2$\sigma$ level) in {\it all} filters bluewards from
the Lyman limit. In this respect, CFHTLS-D3 data are particularly useful to
provide robust non-detection constraints for candidates at z$\sim [$4.5 -- 7$]$. 
Detailed selection criteria are provided below. 
In the subsequent sections we compare the samples
selected and the global properties derived when using different approaches. 

\begin{figure}
   \centering
   \includegraphics[width=0.48\textwidth]{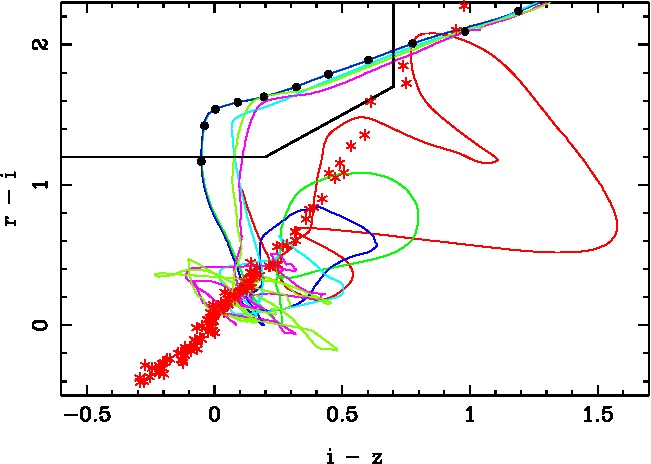}
   \includegraphics[width=0.48\textwidth]{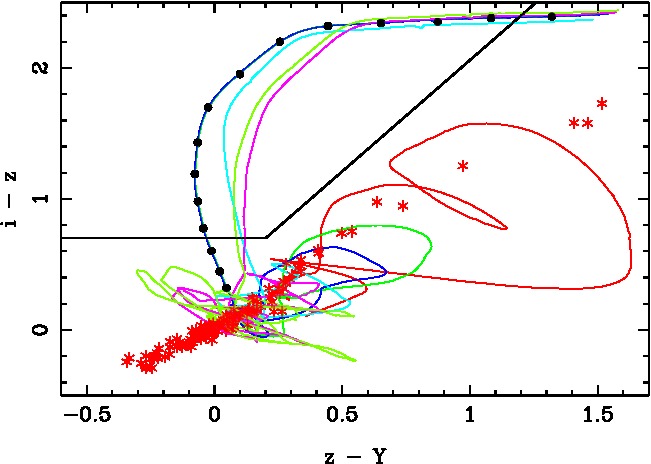}
   \includegraphics[width=0.48\textwidth]{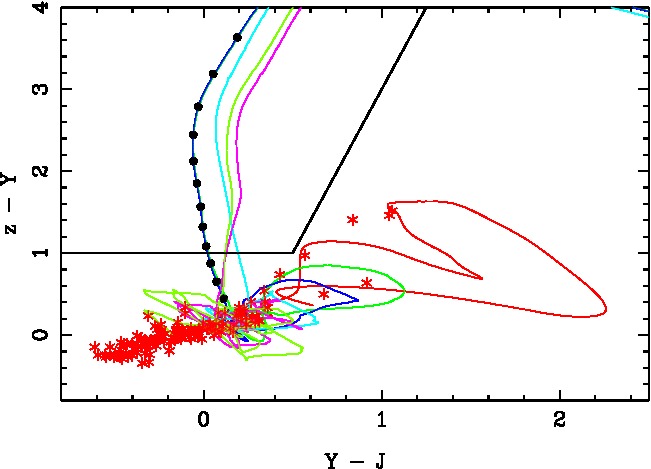}
      \caption{Color-color diagrams showing the position expected for spectral
templates with redshifts $z \sim$ 0--7.0: E-type galaxies (CWW; red solid
line), Scd (CWW; blue lines), Im (CWW; cyan lines),
and starburst templates of \cite{Kinney1996} (magenta and green lines).
Red stars show the expected colors of typical stars based on the
Pickles library \citep{Pickles1998}. 
Black lines delimit the selection windows for the $riz$ 
(z$\sim [$4.5 -- 5.3$]$; top panel), $izY$
(z$\sim [$5.3 -- 6.4$]$; mid panel), and $zYJ$
(z$\sim [$6.3 -- 7.2$]$; bottom panel) dropouts. 
To guide the eye, black dots indicate the redshifts from 
$z=$4.5 to 5.5 (top panel), $z=$5.0 to 6.5 (mid panel), and 
$z=$6.0 to 7.2 (bottom panel), with $\Delta z=$0.1, for a typical star-forming
galaxy. 
}
         \label{fig_cc_diagrams}
   \end{figure}

Regarding the dropout technique,
three selection windows have been used to cover the z$\sim [$4.5 -- 7$]$ interval,
as shown in Fig.~\ref{fig_cc_diagrams}: 

\begin{itemize}

\item [(a)] $r-i > 1.2$, $i-z < 0.7$ and $r-i > 1.0(i-z)+1.0$. This window
selects $r$-dropout candidates in the z$\sim [$4.5 -- 5.3$]$ interval, as shown
in the top panel of Fig.~\ref{fig_cc_diagrams}. This selection window is 
analogous to the one used in the literature to isolate RIz LBGs in the same
redshift range from foreground interlopers and galactic stars
\citep[see e.g.,][]{Ouchi2004, Yoshida2006, 2015ApJ...803...34B, 2017arXiv170406004O}.

\item [(b)] $i-z > 0.7$ and $i-z>1.7(z-Y)+0.35$, a window selecting
$i$-dropout candidates in the z$\sim [$5.3 -- 6.4$]$ interval, as shown in the
mid panel of Fig.~\ref{fig_cc_diagrams}. This selection window is analogous to
the one used in other studies previous
\citep[see e.g.,][]{Bouwens2007, 2015ApJ...803...34B}, but
less strict than in the selection conducted by \cite{Willott2013}. 

\item [(c)] $z-Y > 1.0$ and $z-Y > 4(Y-J)-1.0$. This window is intended to
select $z$-band dropouts in the z$\sim [$6.3 -- 7.2$]$ interval, as shown in
the bottom panel of Fig.~\ref{fig_cc_diagrams}. The field of view in this case is
limited to the WUDS region (see Table~\ref{tab_phot} and
Fig.~\ref{WUDS_weight}). The use of deep $Y$-band images is particularly
useful in this redshift interval and it is rarely found in the literature
excepted for HST data \citep[see e.g.,][]{Bouwens2011B, 2015ApJ...803...34B}. 

\end{itemize}

   Photometric redshifts and associated probability distributions have several
advantages with respect to the three-band dropout technique 
\citep[see e.g.,][]{McLure2009, 2015ApJ...810...71F}. Although the
later have proven to be successful in isolating high-z galaxies, a
significant fraction of the whole population could have been excluded due to
different reasons (e.g., older stellar populations, redder colors, ...). SED
fitting analysis is particularly useful for objects lying close to the
boundaries of the color-color selection boxes. The selection criteria in this
case are simply given by magnitude-selected catalogs (at least 5$\sigma$ in
the filter encompassing the rest-frame UV), and a detection below 2$\sigma$
level in all filters bluewards with respect to 912$\AA$ rest-frame. 
We have adopted the redshift probability distributions (hereafter $P(z)$) obtained when
applying the procedure described in Sect.~\ref{photoz}. Given the selection
based on rest-frame UV, the final sample is still expected to be biased
toward star-forming and low-reddening galaxies. 

\begin{table}
\caption{\label{samples_highz} Number of sources included in the different
redshift bins, for the different selection criteria and input catalogs. } 
\centering
\begin{tabular}{lccc}
\hline
                     &         &   C1               & C2      \\
                     & ($i+z$) &  ($Y+J$)           & ($H+K_s$) \\

Criteria             & N       &   N                   &  N    \\
\hline
$r-$dropout (raw)               & 2016 &  2591  &  2797         \\
z$\sim [$4.5 - 5.3$]$ CC window &  863 &  1085  &  1205         \\
CC window corrected             &  817 &   711  &   -          \\
\hline
$i-$dropout (raw)               &    &  166 &  134        \\
$i-$dropout (corrected)         &    &   91 &   71       \\
z$\sim [$5.3 - 6.4$]$ CC window &     &   98 &   66          \\
CC window corrected             &     &   48 &   32           \\
\hline
$z-$dropout (raw)               &     &  212 &   256     \\
$z-$dropout (corrected)         &     &  142 &   132     \\
z$\sim [$6.3 - 7.2$]$ CC window &     &   36 &    30     \\
CC window corrected             &     &   14 &    11    \\
\hline

\hline
\end{tabular}
\end{table}

\subsection{Samples of galaxies at z$\sim [$4.5 -- 7$]$}
\label{samples}

   The results obtained when applying the selection criteria
described in Sect.~\ref{selection} are presented in this section. 
Table ~\ref{samples_highz} summarizes these results for the different redshift
bins and selection criteria. In all cases we compare the samples extracted
from the different detection images and corresponding catalogs, namely $i+z$,
C1 ($J+Y$) and C2 ($H+K_s$), and we restrict the selection area to the region
covered by all filters involved in the detection with at least 50\% of the
total exposure time, excluding noisy areas (e.g., around bright stars). 
The following samples have been selected:
     
\begin{itemize}

\item [(a)] $r-$dropout sample: A S/N higher than 5$\sigma$ is requested in $i$ and $z$, together
with a non-detection at less than 2$\sigma$ level in $u^{\ast}$ and $g$, and $r-i >
1.2$. This blind selection provides $\sim$2600(2800) sources in C1(C2)
respectively, over the $\sim$1200 arcmin$^2$ field covered by near-IR data. 
Among them, $\sim$1100(1200) are included within the color-color window for
the selection of candidates in the z$\sim [$4.5 -- 5.3$]$ interval.    
When using the $i+z$ detection image instead, $\sim$2000 sources are found over
the same area, 863 of them within the color-color window. An average(median) value of
\zphot=4.82(4.80) is found for this sample. Although the number
of sources is smaller in the later case, the $i+z$-selected sample contains a smaller
fraction of false positives, as discussed below. This $i+z$-selected sample
will be used to derive the LF at z$\sim$5. 

\item [(b)] $i-$dropout sample: A S/N higher than 5$\sigma$ is requested in
$z$ and $Y$, together with a non-detection at less than 2$\sigma$ level in
$u^{\ast}$, $g$ and $r$, and $i-z > 0.7$. In this case the selection is only
applied to the deep WUDS region covered by the $Y$ band. 
These criteria blindly select 166(134) sources in C1(C2) over $\sim$390
arcmin$^2$, and among them 98(66) objects included within the color-color
window for the selection of candidates at z$\sim [$5.3 -- 6.4$]$.
An average(median) value of \zphot=5.84(5.89) is found for this sample.

\item [(c)] $z-$dropout sample: A S/N higher than 5$\sigma$ is requested in
$Y$ and $J$, together with a non-detection at less than 2$\sigma$ level in 
$u^{\ast}$, $g$, $r$ and $i$, and $z-Y > 1.0$. As in the previous case, only
the WUDS region is used for this selection. When these criteria are blindly
applied, 655(627) objects are selected in C1(C2). However, there is a difference
between this sample and the previous ones due to the depth in the $z$-band
filter, which is $\sim$0.5 magnitudes shallower than the $i$-band, as shown in Table
~\ref{tab_phot}. A robust $z-dropout$ selection based on $z-Y$ is only achieved
for objects with $Y<25.50$. When introducing this additional constraint, the
final sample reduces to 212(256) objects in C1(C2) over $\sim$390 arcmin$^2$,
and among them only 36(30) objects are included within the color-color window
for the selection of candidates at z$\sim [$6.3 -- 7.2$]$.
An average(median) value of \zphot=6.83(6.87) is found for this sample.

\end{itemize}

   We have also corrected for obvious spurious sources in the above catalogs
by visual inspection carried out by two different observers. 
The percentage of spurious sources in the $r-$dropout sample is smaller than for the $i-$ and
$z-$dropouts, reaching only $\sim$5\% for the $i+z$ detection image. This
trend was somewhat expected because a good S/N was requested in both $i$ and
$z$, that is at least two visible bands, together with a detection on the near-IR
images for C1 and C2, making the selection of spurious signal highly unlikely.
On the contrary, the contamination is expected to be much higher when the
selection is essentially based on near-IR images, with extraction in
double-image mode and a low detection threshold, together with a poor
detection or not-detection in the optical bands. The presence of spurious
sources is indeed larger in this case, reaching between $\sim$30 and 60\% of the sample,
depending on the detection image and selection window. Therefore we have visually
inspected and validated all objects used in the subsequent analysis. Corrected
counts are reported in Table ~\ref{samples_highz} together with the raw counts,
to illustrate this effect. A detailed study of the $z-$dropout sample is presented
in Paper II, in particular the contamination affecting the z$\sim$7 sample.  


\subsection{Luminosity functions at z$\sim$5 and z$\sim$6}
\label{LF}

   In this section we derive the UV LF at 1500 \AA\  in
two different redshift bins around z$\sim$5 and z$\sim$6, based on the
near-IR-selected samples of WUDS reported in Table ~\ref{samples_highz}. 
The results obtained on the LF at z$\ge$7, as well as the evolution of the UV
LF between z$\sim$ 4.5 and 9, are presented and discussed in Paper II.

   For each redshift bin, two different approaches and samples
have been used to derive the LF. In one hand, the complete $r-$dropout and $i-$dropout
catalogs, corrected for spurious detections, without any additional color-selection, are
used to compute the LF based on photometric redshifts probability
distributions ($P(z)$). On the other hand, we have used the (corrected)
subsamples included within the corresponding LBG color-color selection
windows. These two different approaches and samples are widely used in the literature.
 
   Number density values in luminosity bins have been computed using the $1/V_{max}$
method \citep[see e.g.,][]{Schmidt1968}. The bootstrap approach developed by
\cite{Bolzonella2002} has been adopted to compute the LF data points, that is number
density values and confidence intervals. This method is based on blind
photometric redshifts and associated probability distributions $P(z)$.
For a given redshift interval, 1000
realizations of each catalog have been performed; for each realization of the
catalog and for each object, a random value of the photometric redshift was
sorted out according to its $P(z)$. This procedure takes into account by
construction the existence of degenerate solutions in redshift for a given
source. The number of realizations of each catalog is large enough to ensure
that the LF results do not depend on the number of realizations. 

   Absolute magnitudes in the UV at $\sim$1500\AA \ (M$_{1500}$) are derived from
the photometric SED datapoints overlapping this wavelength for a given
(photometric) redshift. We have checked that there is no significant
difference in the LF results when using the closest filter-band, assuming a
flat continuum, or the precise flux at rest-frame 1500\AA \ for the best-fit
model instead. 

   Number densities have been corrected for photometric incompleteness depending on
the selection bands, according to Sect. ~\ref{completeness}. For samples
selected in LBG windows, an additional multiplicative correction
was applied to include the effect of color selection as a function of
redshift and magnitude (S/N ratio) in the detection filter encompassing the
1500\AA \ rest-frame. The shape of this
later correction was obtained through simulated catalogs, each one containing
$\sim$4 $\times$ 10$^{5}$ objects, fully covering the redshift windows
$[$4.5,5.5$]$ and $[$5.5,6.5$]$. These catalogs are based on the same
templates used to define the selection windows, 
with magnitudes and corresponding photometric errors sorted to uniformly sample
the actual range of magnitudes in the WUDS survey. We have then applied to these
simulated catalogs the same color selection as for real data. The correction
factor in a given redshift and magnitude bin is simply derived as the ratio
between the number of galaxies in the input sample and number actually
retrieved by the selection process. Also the redshift interval defined by
$r-$dropout and $i-$dropout selections is actually narrower than the nominal
$\Delta$z=1. In order to facilitate the comparison between the different
samples, we have used a fixed $\Delta$z=1 in all cases assuming a uniform number
density of sources. 
Therefore, in the following, number densities are given for redshift bins $[$4.5,5.5$]$
and $[$5.5,6.5$]$ respectively for convenience, and also to facilitate the
comparison with other surveys. 


   LF values have been computed using regularly spaced bins in luminosity
of $\Delta M_{1500}$=$\pm$0.125 at z$=$5 and $\Delta M_{1500}$=$\pm$0.250 at z$=$6,
excepted for the first (brightest) bins, arbitrarily set to $\Delta M_{1500}$=0.25
mags or 0.50 to improve statistics. Note that the number of sources used in
the blind photometric redshift approach is larger than in the LBG color-color
window (because less restrictive; see Table ~\ref{samples_highz}),
allowing us to better sample the LF at z$=$6. 
The fact that no source was detected at magnitudes brighter
than M$_{1500}\le$ $-22.4$ and $-21.9$ respectively for $r$ and $i-$dropouts
has been used to determine upper limits for the bright end of the LF. 

    It should be noted that results obtained at z$=$5 strongly depend on the
detection image used for the $r-$dropout selection.
In the following and for a sake of consistency, we
use the catalogs derived from detection images encompassing the rest-frame
1500\AA \ (see discussion below), that is $i+z$ at z$=$5 and C1 at z$=$6. 
Tables ~\ref{LF_results5} and ~\ref{LF_results6} summarize the LF data
points at z$=$5 and z$=$6 respectively, for the different detection
catalogs. In all cases, error bars corresponding to 68\% (1$\sigma$)
confidence levels resulting from the bootstrap process are combined together with
an estimate of Poisson uncertainties and field-to-field variance
derived from the public cosmic variance calculator by
\cite{Trenti-Stiavelli2008}. These later contributions clearly dominate the
error budget, excepted for the brightest M$_{1500}\le$ $-22$ luminosity bins.

Tables ~\ref{LF_results5} and ~\ref{LF_results6} present the LF points
derived at z$=$5 and z$=$6 using the two approaches/samples to compute the LF,
namely photometric redshifts applied to blindly selected dropout samples,
and the classical LBG color-preselected samples. Results are found to be
consistent, in general at better than $\sim$1 $\sigma$ level, the largest
deviations being observed for the faintest luminosities at z$=$6. 
As discussed below, these results are also in agreement
with previous findings at the same redshifts, as if photometric
redshifts and associated probability distributions could be safely
used for these purposes on the dropout samples, without introducing
more sophisticated and to a certain extent dangerous cuts in color space. 
This statement requires some additional discussion (see below).


   \begin{figure}
   \centering
   \includegraphics[width=0.48\textwidth]{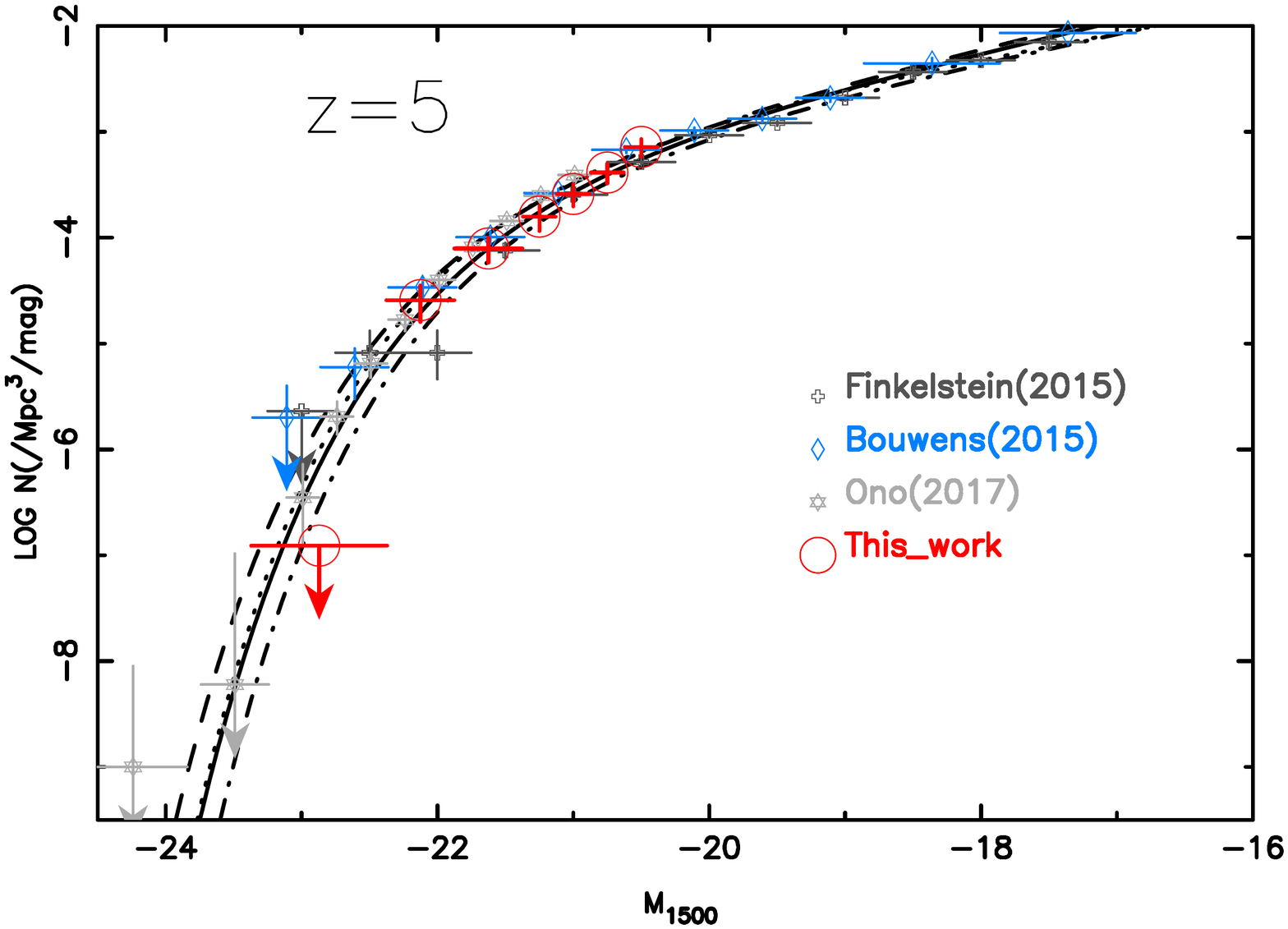}
      \caption{Comparison of the WUDS luminosity function at z$=$5 
(based on $i+z$ detection) with equivalent estimates by
\cite{2015ApJ...810...71F} (black crosses, dot-dashed line), 
\cite{2015ApJ...803...34B} (blue diamonds, dashed line), and 
\cite{2018PASJ...70S..10O} (gray stars, dotted line). 
The best Schechter fit to WUDS (LBG color selection) + extended data is plotted by a solid line. 
}
         \label{LF5_plot}
   \end{figure}

   \begin{figure}
   \centering
   \includegraphics[width=0.45\textwidth]{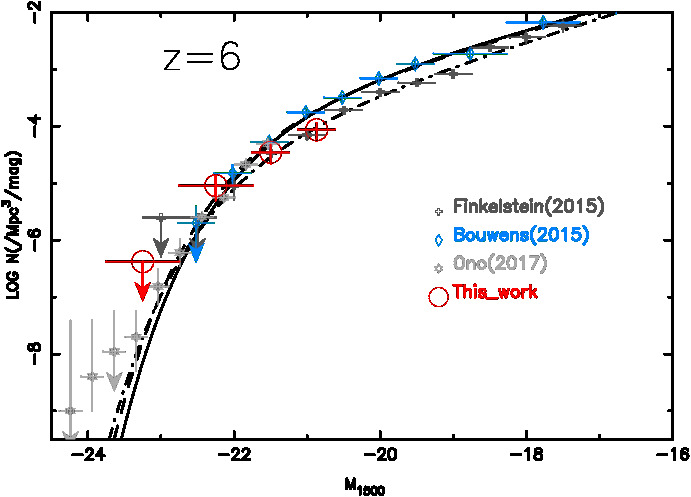}\\
   \includegraphics[width=0.45\textwidth]{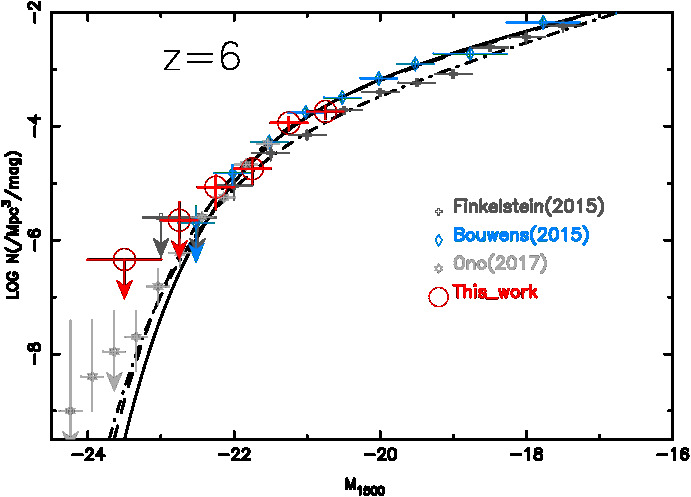}
   \caption{Comparison of the WUDS luminosity function at z$=$6 with the
equivalent estimates by
\cite{2015ApJ...810...71F} (black crosses, dot-dashed line), 
\cite{2015ApJ...803...34B} (blue diamonds, dashed line), and 
\cite{2018PASJ...70S..10O} (gray stars, dotted line). 
for the LBG color-selected sample (top) and photometric redshifts applied to the $i-$drop
sample (bottom). The best Schechter fits to WUDS + extended data are plotted by solid lines.
For the LBG color-selected sample presented in this figure,
the slope has been fixed as in \cite{2015ApJ...803...34B} 
(see Sect.\ \ref{LF}). 
}
         \label{LF6_plot}
   \end{figure}


   A subsequent maximum-likelihood fit of data points has been performed to the
analytic Schechter function \citep{1976ApJ...203..297S} as follows:
\begin{equation}
$$
\Phi(M) dM =
\Phi^{\star}_{1500} \frac{ln(10)}{2.5} 
\left(10^{-0.4(M-M^*)}\right)^{\alpha+1} 
\exp \left(-10^{-0.4(M-M^*)}\right) dM
$$
\end{equation}
\noindent
based on a simple least-squares $\chi^2$ minimization,   
assuming that the Schechter function provides a good representation of data.
The fit has been obtained separately for data points derived
through the classical LBG color-selection window and blind photometric redshifts
to facilitate the comparison with previous findings. 

Our data being essentially sensitive to the normalization and M$^{\star}$,
we have studied the influence of the slope $\alpha$ in two different ways: by adding
data points from the literature toward the faintest luminosities, and by
setting the slope of the LF. 
First, we completed our step-wise data points toward the faint edge using the LF estimates 
by \cite{2015ApJ...803...34B} at z$=$5 and z$=$6, based on deep HST imaging.
A good-quality fit was obtained at z$=$5, with stable and consistent results for both the
LBG color-selection window and blind photometric redshifts, also consistent with the
slope value derived by \cite{2015ApJ...803...34B}. 
On the contrary, at z$=$6 a good-quality fit could not be achieved for the enlarged data
set based on the LBG color-selection window, but only for the photometric redshift LF data  
(for which the sampling of the LF is better, as shown in Table~\ref{LF_results6}). 
Therefore, in a second step, we imposed a constant slope for the LF following \cite{2015ApJ...803...34B}
(also based on LBG color window) while leaving the normalization and M$^{\star}$ free.
In this case, a good fit was also achieved at z$=$6, yielding consistent results
for both the LBG color-selection window and blind photometric redshifts.
The effect of imposing a constant slope for the LF was studied by using also the
steeper value from the \cite{2016PASA...33...37F} review 
and from \cite{2017ApJ...835..113L} in lensing fields, 
both studies being based on photometric redshifts.
As seen in Table ~\ref{LF_comparison}, our results for $\Phi^{\star}$ and L$^{\star}$ are
in good agreement with those previously found by \cite{2016PASA...33...37F}
and \cite{2017ApJ...835..113L} when imposing the same value of $\alpha$ (see below). 

Our best fit to WUDS data-sets yields
the results summarized in Table ~\ref{LF_comparison}, namely 
$\Phi^{\star}_{1500}=$(8.20$^{+0.3}_{-0.5}$)$\times$10$^{-4}$ Mpc$^{-3}$,
M$^{\star}=$-20.98$\pm$0.03 mag, and 
$\alpha=$-1.74 $\pm$0.02 for z$=$5, and 
$\Phi^{\star}_{1500}=$(7.5$^{+2.5}_{-2.5}$)$\times$10$^{-4}$ Mpc$^{-3}$ and 
M$^{\star}=$-20.58$\pm$0.022 mag with fixed  
$\alpha=$-1.87 for z=6,  
for the LBG color-selection window. 
Figure\ \ref{LF5_plot} and \ref{LF6_plot} display the data points 
adopted for the LBG color-selection window for z$\sim$5 and z$\sim$6
respectively, together with the best-fit representation by a
Schechter function as discussed above.
Our results confirm a slight evolution in the UV LF between z$=$5 and z$=$6, 
consistent with a dimming of both $\Phi^{\star}$ and L$^{\star}$. 

Oddly, when comparing our results at z$\sim$6 with previous findings in Fig.\ \ref{LF6_plot}, 
at first glance they are found to be more consistent with \cite{2015ApJ...803...34B} 
for the blind photometric redshift approach, whereas they are more consistent with
\cite{2015ApJ...810...71F} 
for the LBG color-selection window, while the opposite trend would have been expected.
In reality, as seen in Table ~\ref{LF_comparison}, the results of the Schechter fit
derived from both the color-selection window and the photometric redshift
approaches are fully consistent within the error bars with \cite{2015ApJ...803...34B} when
imposing the same slope $\alpha=$-1.87, the larger differences being in the normalization.
When imposing a steeper slope $\alpha$ $\sim$-2, as found by
\cite{2016PASA...33...37F} or 
\cite{2017ApJ...835..113L} based on the photometric redshift approach, 
then the results of the fit are fully consistent with their findings regarding
$\Phi^{\star}$ and L$^{\star}$. In other words, the results obtained on L$^{\star}$ and the
normalization seem to be more sensitive to the imposed value for the slope $\alpha$
rather than to the approach/method used to select the samples. 

It is worth to mention that the difference between the LF points derived from the
two samples/approaches defined within the same WUDS field provides an estimate of the
systematic uncertainties related to the selection function, that are not necessarily
taken into account in the literature \citep[see ][for a review]{2016PASA...33...37F},
meaning that the final errors on the LF points and fits could be systematically underestimated.
In this study, the LF points derived at z$\sim$5 are fully consistent within the error bars
between the two approaches, making the LF parameters particularly robust. At z$\sim$6,
the difference between the two approaches is negligible for the brightest region, that is
smaller than $\sim$30\% up to M$_{1500}\le$ -21.5 (similar or smaller than the error bars),
whereas at lower luminosities the photometric redshift approach yields
$\Delta log \Phi \sim$ 0.3 higher in average. When quadratically combining these
differences with the current error bars, as if they were the expression of a systematic error,
we find that the fit results for the LF are very slightly modified (e.g., $\sim$0.01 for the slope when it is let free, $\sim$0.02 for M$_{1500}$, and a negligible amount for the normalization, the
error bars on these parameters increasing by only $\sim$10\%). This means that there is indeed
a systematic effect to take into account, but it is not strong enough here to modify the
conclusions of this study.  
  
\begin{table*}
\caption{\label{LF_results5}Luminosity function at z$=$5 in the WUDS field }
\centering
\begin{tabular}{ccc|ccc}
\hline
\multicolumn{3}{c}{$r-$drop Photometric Redshifts} &
\multicolumn{3}{c}{LBG color window} \\
\hline
M$_{1500}$ & $\Phi$ & $\Delta \Phi$ &
M$_{1500}$ & $\Phi$ & $\Delta \Phi$ \\

[AB mag] & [Mpc$^{-3}$ mag$^{-1}$] & [Mpc$^{-3}$ mag$^{-1}$] &
[AB mag] & [Mpc$^{-3}$ mag$^{-1}$] & [Mpc$^{-3}$ mag$^{-1}$] \\
\hline
-22.875$\pm$0.500  & $<$1.23 10$^{-7}$  &       &
-22.875$\pm$0.500  & $<$1.23 10$^{-7}$  &        \\
-22.125 $\pm$0.250 & 1.86 10$^{-5}$ & 0.77 10$^{-5}$  &
-22.125 $\pm$0.250 & 2.57 10$^{-5}$  &   0.97 10$^{-5}$ \\
-21.625 $\pm$0.250 & 5.86 10$^{-5}$ & 2.12 10$^{-5}$  &
-21.625 $\pm$0.250 & 7.90 10$^{-5}$ & 2.10 10$^{-5}$    \\ 
-21.250 $\pm$0.125 & 1.73 10$^{-4}$ & 0.44 10$^{-4}$  &
-21.250 $\pm$0.125 & 1.58 10$^{-4}$ & 0.42 10$^{-4}$   \\
-21.000 $\pm$0.125 & 2.71 10$^{-4}$ & 0.68 10$^{-4}$  &
-21.000 $\pm$0.125 & 2.59 10$^{-4}$ & 0.61 10$^{-4}$    \\
-20.750 $\pm$0.125 & 7.86 10$^{-4}$ & 1.60 10$^{-4}$  &
-20.750 $\pm$0.125 & 4.13 10$^{-4}$ & 0.87 10$^{-4}$    \\ 
-20.500 $\pm$0.125 & 9.51 10$^{-4}$ & 1.94 10$^{-4}$  &
-20.500 $\pm$0.125 & 7.19 10$^{-4}$ & 1.36 10$^{-4}$  \\
\hline
\end{tabular}
\tablefoot{$\Delta \Phi$ includes the 68\% (1$\sigma$) confidence level
  intervals from the bootstrap procedure, Poisson uncertainties and
  field-to-field variance. 
}
\end{table*}


\begin{table*}
\caption{\label{LF_results6}Luminosity function at z$=$6 in the WUDS field}
\centering
\begin{tabular}{ccc|ccc}
\hline
\multicolumn{3}{c}{$i-$drop Photometric Redshifts} &
\multicolumn{3}{c}{LBG color window} \\
\hline
M$_{1500}$ & $\Phi$ & $\Delta \Phi$ &
M$_{1500}$ & $\Phi$ & $\Delta \Phi$ \\

[AB mag] & [Mpc$^{-3}$ mag$^{-1}$] & [Mpc$^{-3}$ mag$^{-1}$] & 
[AB mag] & [Mpc$^{-3}$ mag$^{-1}$] & [Mpc$^{-3}$ mag$^{-1}$] \\
\hline
-22.375 $\pm$0.50  & $<$4.19 10$^{-7}$  &       & 
-22.375 $\pm$0.50  & $<$4.19 10$^{-7}$  &      \\ 
-22.75  $\pm$0.25  & 2.25 10$^{-6}$  & 2.41 10$^{-6}$  &
-22.25  $\pm$0.50  & 9.07 10$^{-6}$  & 4.0 10$^{-6}$ \\
-22.25  $\pm$0.25  & 8.56 10$^{-6}$  & 5.08 10$^{-6}$  &
-21.50  $\pm$0.25  & 3.51 10$^{-5}$ & 1.24 10$^{-5}$  \\
-21.75  $\pm$0.25  & 1.80 10$^{-5}$ & 0.82 10$^{-5}$  &
-20.87  $\pm$0.25  & 8.72 10$^{-5}$ & 2.62 10$^{-5}$  \\
-21.25  $\pm$0.25  & 1.15 10$^{-4}$ & 0.30 10$^{-4}$ &
                   &               &         \\
-20.75  $\pm$0.25  & 1.82 10$^{-4}$ & 0.47 10$^{-4}$ &
                   &               &         \\
\hline
\end{tabular}
\tablefoot{$\Delta \Phi$ includes the 68\% (1$\sigma$) confidence level
  intervals from the bootstrap procedure, Poisson uncertainties and
  field-to-field variance. 
}
\end{table*}

\subsection{Discussion. Comparison with previous findings}
\label{comparison}

   The LF at z$=$5 and z$=$6 has been the subject of various studies during the last
ten years 
\citep[see e.g.,][]{Yoshida2006,Bouwens2007,McLure2009,Su2011,
Bouwens2012B,Willott2013,2015ApJ...810...71F,2015ApJ...803...34B,
2016PASA...33...37F,2017ApJ...835..113L,2017arXiv170406004O}.
Although detailed results on the evolution of the LF will be presented in
Paper II, the comparison between the present results and previous findings is
important to assess the quality of the survey. WUDS covers an interesting region in
luminosity between the exponential and power-law-dominated regimes of the LF.  
WUDS has been designed to provide constraints on the brightest part of the LF at
high-z, in particular on M$^{\star}$ and $\Phi^{\star}$, and it is much less
sensitive to the value of $\alpha$, as discussed in Sect.~\ref{LF}. 
We compare in Table ~\ref{LF_comparison} the
LF parameters obtained in this survey at z$=$5 and z$=$6 with current findings
in the literature, in complement to Fig.\ ~\ref{LF5_plot} and
\ref{LF6_plot}. 

As expected by construction, the best-fit values when the slope $\alpha$ is let free
are fully consistent with \cite{2015ApJ...803...34B} because we used their
data points to complete the LF toward the faintest luminosities. It should be mentioned however
that, at z$=$6, our faintest bins are in agreement with \cite{2015ApJ...803...34B} for the
photometric redshift ``blind'' approach, whereas a better agreement is found with
\cite{2015ApJ...810...71F} when using the LBG color-window, as seen in
Fig.\ ~\ref{LF6_plot},
while the opposite trend would have been expected.
As discussed in Sect.~\ref{LF}, given the error bars, the results obtained on
$\Phi^{\star}$ and L$^{\star}$ seem to be more sensitive to the value imposed for the
slope $\alpha$ rather than to the method used to select the samples. 

With the clarifications given above,
our results are consistent with a small evolution of both
M$^{\star}$ and $\Phi^{\star}$ between z$=$5 and z$=$6. 
There is also a good agreement in general between our M$^{\star}$
and $\Phi^{\star}_{1500}$ values and recently published results within the
error bars. At z$=$6 this is true provided that the same value for the slope $\alpha$ is imposed.
At z$=$5, the present combined determination of M$^{\star}$ and  $\Phi^{\star}$ is more
accurate then previous findings due to a better statistics in the intermediate-luminosity
regime. Regarding the value of the slope at this redshift, we confirm the value proposed by \cite{2015ApJ...803...34B},
and we are marginally consistent with \cite{2015ApJ...810...71F} and \cite{2017arXiv170406004O}.
Our results lend support to higher $\Phi^{\star}_{1500}$ determinations than usually reported
at z$=$6, but still consistent with the latest findings by \cite{2015ApJ...803...34B},
\cite{2016PASA...33...37F}, and \cite{2017arXiv170406004O}.
Of particular interest is the comparison with
\cite{Willott2013} at z$=$6 (Table 3 in their paper), because it was based on
CFHTLS-Deep data partly covering the WIRDS area. Even though the selection
criteria are not the identical (see Sect.~\ref{selection}), and their LF was
obtained at 1350 instead of 1500 \AA \ , the global fit to our data as well as
the number densities obtained in the present paper are consistent with their
results. 
Given the error bars, our results are consistent with an evolution
in L$^{\star}$, as expected if the UV luminosity of galaxies follows the assembly of
host dark matter halos \citep[see e.g.,][]{Bouwens08}, without excluding an evolution on the
global normalization, as reported for instance by \cite{2015ApJ...803...34B}. 
A direct comparison is recognized to be difficult given the degeneracy between the Schechter
parameters on one hand, and the differences in the methods and samples used to build the
LF on the other hand.

\begin{table*}
\caption{\label{LF_comparison} Comparison between the LF parameters at z$=$5
  and z$=$6 in the recent literature}
\centering
\begin{tabular}{lccc}
\hline
\multicolumn{4}{c}{z$=$5} \\
\hline
References & $\alpha$ & M$^{\star}$ & $\Phi^{\star}_{1500}$  \\
           &           &   [FUV AB mag] & [10$^{-4}$ Mpc$^{-3}$] \\
\hline
{\bf WUDS (LBG color window)} & {\bf -1.74 $\pm$0.02} & {\bf -20.98$\pm$0.03} & {\bf 8.20$^{+0.3}_{-0.5}$} \\
{\bf WUDS (Photometric Redshifts)} & {\bf -1.74 $\pm$0.04} & {\bf -21.02$\pm$0.08} & {\bf 8.00$^{+0.8}_{-0.6}$} \\
\cite{McLure2009}  & -1.66 $\pm$0.06 & -20.73$\pm$0.11 & 9.4$\pm$0.19 \\
\cite{2015ApJ...810...71F} & -1.67$\pm$0.06 & -20.81$^{+0.13}_{-0.12}$ & 8.95$^{+1.92}_{-1.31}$ \\  
\cite{2015ApJ...803...34B} & -1.76$\pm$0.06 & -21.17$\pm$0.12 & 7.4$^{+1.8}_{-1.4}$ \\ 
\cite{2018PASJ...70S..10O} & -1.60$\pm$0.06 & -20.95$\pm$0.06 & 10.7$^{+1.3}_{-1.1}$ \\  
\hline
\multicolumn{4}{c}{z$=$6} \\
\hline
References & $\alpha$ & M$^{\star}$ & $\Phi^{\star}_{1500}$  \\
           &           &   [FUV AB mag] & [10$^{-4}$ Mpc$^{-3}$] \\
\hline
{\bf WUDS (LBG color window)}      & {\bf -1.87 (fixed)}    & {\bf -20.58$\pm$0.22} & {\bf 7.5$^{+2.5}_{-2.5}$} \\
{\bf WUDS (Photometric Redshifts)} & {\bf -1.84$\pm$0.09} & {\bf -20.77$\pm$0.20} & {\bf 6.5$^{+2.0}_{-1.6}$} \\
{\bf WUDS (Photometric Redshifts)} & {\bf -1.87 (fixed)} & {\bf -20.83$\pm$0.22} & {\bf 5.8$^{+1.8}_{-1.4}$} \\
{\bf WUDS (Photometric Redshifts)} & {\bf -1.91 (fixed)} & {\bf -20.80$\pm$0.30} & {\bf 5.7$^{+2.6}_{-2.1}$} \\
{\bf WUDS (Photometric Redshifts)} & {\bf -2.10 (fixed)} & {\bf -21.20$^{+0.10}_{-0.15}$ } & {\bf 2.45$^{+0.59}_{-0.40}$} \\
\cite{McLure2009}  & -1.71$\pm$0.11  & -20.04$\pm$0.12 & 1.80$\pm$0.50 \\
\cite{Su2011}      & -1.87$\pm$0.14  & -20.25$\pm$0.23 & 1.77$^{+0.62}_{-0.49}$ \\
\cite{2015ApJ...810...71F} & -2.02$\pm$0.10 & -21.13$^{+0.25}_{-0.31}$ & 1.86$^{+0.94}_{-0.80}$ \\    
\cite{2015ApJ...803...34B} & -1.87$\pm$0.10 & -20.94$\pm$0.20 & 5.0$^{+2.2}_{-1.6}$ \\ 
\cite{2015MNRAS.452.1817B} & -1.88$^{+0.15}_{-0.14}$ & -20.77$^{+0.18}_{-0.19}$ & 5.7$^{+2.7}_{-2.0}$ \\ 
\cite{2016PASA...33...37F} & -1.91$^{+0.04}_{-0.03}$ & -20.79$^{+0.05}_{-0.04}$ & 4.26$^{+0.52}_{-0.38}$ \\ 
\cite{2017ApJ...835..113L} & -2.10$\pm$0.03 & -20.826$^{+0.051}_{-0.040}$ &  2.254$^{+0.20}_{-0.16}$ \\ 
\cite{2018PASJ...70S..10O} & -1.86$\pm$0.07 & -20.90$\pm$0.07 & 5.5$^{+0.9}_{-0.9}$ \\ 
\hline
\end{tabular}
\end{table*}
\section{Conclusions and Perspectives}
\label{conclusion}

   In this paper we have introduced and characterized the WIRCam Ultra Deep Survey
(WUDS), a 4-band near-IR photometric survey covering $\sim$400 arcmin$^2$ on
the CFHTLS-D3 field (Groth Strip). This public survey was specifically
tailored to set strong constraints on the cosmic SFR and the UV 
luminosity function brighter or around $L^{\star}$ in the z$\sim$6-10 domain.

Regarding the properties of the data at low-$z$,
and according to the estimates presented in this article, WUDS should allow the
users to detect
early-type galaxies with stellar masses down to log(M$^{\star}$/M$\odot$)$>$10.5$\pm$1.0,
and late-type galaxies down to log(M$^{\star}$/M$\odot$)$>$9.9$\pm$0.60 up to z$\sim$3.5.
The quality of the photometric redshifts achieved for the WUDS survey is
comparable to the one obtained by other large surveys when using a similar number of filters
and a similar depth. 

As part of this effort, we have focused in this article on the selection
of galaxy samples at z$\sim [$4.5 -- 7$]$ and the determination of the
UV LF at z$=$5 and z$=$6,
taking advantage from the deep optical data available from the CFHTLS-Deep
Survey. We have also extended the research to an adjacent shallower area of $\sim$1000
arcmin$^2$ extracted from the WIRDS Survey, and observed in three near-IR bands.
Using two different approaches, the classical LBG color-selection technique and
photometric redshifts blindly applied to dropout samples,
we have selected high-z galaxies and computed the UV LF.
At z$=$5, the combined determination of
M$^{\star}$ and $\Phi^{\star}$ is more accurate then previous results due to
a better statistics in the intermediate-luminosity regime.
The evolution in the UV LF between z$=$5 and z$=$6 is consistent with a
small dimming in both $\Phi^{\star}$ and L$^{\star}$.
These results are consistent with
previous findings in terms of the M$^{\star}$ and $\Phi^{\star}$ values, knowing
that WUDS covers a particularly interesting interval at intermediate luminosities.


   The selection and combined analysis of different galaxy samples at z$\ge$7
will be presented in a forthcoming paper, as well as the evolution of the UV
luminosity function between z$\sim$ 4.5 and 9
(Laporte et al., submitted to A \& A, Paper II).
Photometric data and catalogs will be set publicly available at
the \url{http://wuds.irap.omp.eu/} web site.

   WUDS is intended to provide a robust database in the near-IR for the
selection of targets for the Galaxy Origins and Young Assembly (GOYA) Survey.
GOYA is a scientific program to be developed mainly using the guaranteed time of
the international consortium building EMIR, a wide-field, near-IR spectrograph
currently installed in the Nasmyth focus of the Spanish 10.4m GTC at
Canary Islands\footnote{\tt{http://www.gtc.iac.es}} 
\citep{Garzon06}. The GOYA project addresses the formation and evolution of
galaxies, in particular the structure, dynamics and integrated stellar
populations of galaxies at high redshift 
\citep[see e.g.,][]{Guzman03,Balcells03,Balcells07,Dominguez-Palmero08}.  

\begin{acknowledgements}
   Part of this work was supported by the French CNRS, the French {\it Programme
National de Cosmologie et Galaxies} (PNCG) of CNRS/INSU with INP and IN2P3, 
co-funded by CEA and CNES.  
We acknowledge support from the Spanish Programa Nacional de
Astronom\'ia y Astrof\'isica under grant AYA 2006-02358 and the Swiss National Science
Foundation. Partially funded by the
Spanish MEC under the Consolider-Ingenio 2010 Program grant CSD2006-00070:
First Science with the GTC (http://www.iac.es/consolider-ingenio-gtc/).
This work recieved support from Agence Nationale de la recherche bearing the
reference ANR-09-BLAN-0234, and from the  ECOS SUD Program C16U02.
This work has been carried out thanks to the support of the OCEVU Labex
(ANR-11-LABX-0060) and the A*MIDEX project (ANR-11-IDEX-0001-02) funded by the
"Investissements d'Avenir" French government program managed by the ANR.
NL acknowledges financial support from European Research Council Advanced Grant FP7/669253. 
JG and NC acknowledge financial support from Spanish MINECO (AYA2016-75808-R).
MB acknowledges support from AYA grant 2009-11139. 
This paper is based on observations obtained with
WIRCam, a joint project of CFHT, Taiwan, Korea, Canada, France, and the
Canada-France-Hawaii Telescope (CFHT) which is 
operated by the National Research Council (NRC) of Canada, the Institute
National des Sciences de l'Univers of the Centre National de la Recherche
Scientifique of France, and the University of Hawaii.
Funding for the DEEP2 survey has been provided by NSF grants AST95-09298,
AST-0071048, AST-0071198, AST-0507428, and AST-0507483 as well as NASA LTSA
grant NNG04GC89G. 
\end{acknowledgements}

%
\bibliographystyle{aa} 
\bibliography{WUDS_paper1} 
%

\end{document}